\documentstyle[11pt,epsf]{article}

\newcommand{\beq}{\begin{equation}} \newcommand{\eeq}{\end{equation}}
\newcommand{\bea}{\begin{eqnarray}} \newcommand{\eea}{\end{eqnarray}}

  \newcommand
{\Romannumeral}[1]{\uppercase\expandafter{\romannumeral#1}}

\newcommand{\be}{\begin{enumerate}} \newcommand{\ee}{\end{enumerate}}
\newcommand{\bi}{\begin{itemize}} \newcommand{\ei}{\end{itemize}}
\newcommand{\ba}{\begin{array}} \newcommand{\ea}{\end{array}}
\newcommand{\bc}{\begin{center}} \newcommand{\ec}{\end{center}}
\newcommand{\bt}{\begin{tabular}} \newcommand{\et}{\end{tabular}}

\def\lsim{\mathrel{\rlap{\lower4pt\hbox{\hskip1pt$\sim$}}
    \raise1pt\hbox{$<$}}}           
\def\gsim{\mathrel{\rlap{\lower4pt\hbox{\hskip1pt$\sim$}}
    \raise1pt\hbox{$>$}}}           

\newcommand{\Tr}{\mathop{\rm Tr}}           
\newcommand{\tr}{\mathop{\rm tr}}           
\newcommand{\half}{\textstyle {1\over2} \displaystyle}    
\newcommand{\third}{\textstyle {1\over3} \displaystyle}   
\newcommand{\quarter}{\textstyle {1\over4} \displaystyle} 
\newcommand{\sixth}{\textstyle {1\over6} \displaystyle}   
\newcommand{\twoth}{\textstyle {2\over3} \displaystyle}   
\newcommand{\Dslash}{{\hbox{D}\kern-0.6em\raise0.15ex\hbox{/}}} 

\renewcommand{\et}{\eta}

\newcommand{\deltaslash}{\not{\hbox{\kern-2pt $\delta$}}} 

\begin{document}

\setlength{\oddsidemargin}{0cm} \setlength{\baselineskip}{7mm}

\input epsf

\begin{normalsize}\begin{flushright}

 
December 2008 \\

\end{flushright}\end{normalsize}

\begin{center}
  
\vspace{5pt}

{\Large \bf Quantum Gravity on the Lattice} \\

\vspace{40pt}
 
{\sl Herbert W. Hamber}
$^{}$\footnote{Invited lecture presented at the conference "Quantum Gravity: Challenges and Perspectives", Bad Honnef, April 14-16 2008. To appear in the proceedings edited by Hermann Nicolai.}
\vspace{30pt}

Department of Physics \\
University of California \\
Irvine, Ca 92717, USA \\

\vspace{20pt}

\end{center}

\begin{center} {\bf ABSTRACT } \end{center}

\noindent

I review the lattice approach to quantum gravity, and how it relates
to the non-trivial ultraviolet fixed point scenario of the continuum
theory.
After a brief introduction covering the general problem of ultraviolet 
divergences in gravity and other non-renormalizable theories,
I cover the general methods and goals of the lattice approach.
An underlying theme is the attempt at establishing connections
between the continuum renormalization group results, which are 
mainly based on diagrammatic perturbation theory, 
and the recent lattice results, which apply to the strong
gravity regime and are inherently non-perturbative.
A second theme in this review is the ever-present natural correspondence 
between infrared methods of strongly coupled non-abelian gauge theories 
on the one hand, and the low energy approach to quantum gravity based
on the renormalization group and universality of critical behavior 
on the other.
Towards the end of the review I discuss possible observational consequences of 
path integral quantum gravity, as derived from the non-trivial ultraviolet
fixed point scenario.
I argue that the theoretical framework naturally leads to considering 
a weakly scale-dependent Newton's costant, with a scaling violation parameter
related to the observed scaled cosmological constant 
(and not, as naively expected, to the Planck length).





\vfill

\pagestyle{empty}

\newpage

\pagestyle{plain}

\vskip 20pt

\section{Introduction and Motivation}
\label{sec:intro}

Since the seventies strategies that deal with the problem of 
ultraviolet divergences in quantum gravity have themselves diverged.
Some have advocated the search for a new theory of
quantum gravity, a theory which does not suffer from ultraviolet
infinity problems.
In supersymmetric theories, such as supergravity and 
ten-dimensional superstrings, 
new and yet unobserved particles are introduced thus reducing
the divergence properties of Feynman amplitudes.
In other,  very restricted classes of 
supergravity theories in four dimensions, proponents have
claimed that enough
conspiracies might arise thereby making these models finite.
For superstrings, which live in a ten-dimensional spacetime,
one major obstacle prevails to date:
what dynamical mechanism would
drive the compactification of spacetime from the ten dimensional
string universe to our physical four-dimensional world?

A second approach to quantum gravity
has endeavored to pursue new avenues to quantization,
by introducing new quantum variables
and new cutoffs, which involve
quantum Hamiltonian methods based on parallel
transport loops, spacetime spin foam and new types
of quantum variables describing a quantum dust.
It is characteristic of these methods that the underlying
theory is preserved: it essentially remains a quantum 
version of Einstein's relativistic theory, yet the ideas
employed are intended to go past the perturbative treatment.
While some of these innovative tools have had limited success
in exploring the much simpler non-perturbative features of 
ordinary gauge theories, 
proponents of such methods have argued that gravity is fundamentally
different, thereby necessitating the use of radically new methods.

The third approach to quantum gravity, which forms the 
topic of this review, focuses instead on the 
application of modern methods of quantum field theory.
Its cornerstones include
the manifestly covariant Feynman path integral approach,
Wilson's modern renormalization group ideas and the development
of lattice methods to define a regularized form of the path integral,
which would then allow controlled non-perturbative calculations.
In non-abelian gauge theories and in the standard model
of elementary particle interactions, these methods are invariably
the tools of choice:
the covariant Feynman path integral approach is
crucial in proving the renormalizability of non-abelian
gauge theories;
modern renormalization group methods establish the core 
of the derivation of the asymptotic
freedom result and related discussions of momentum dependence
of amplitudes in terms of a running coupling constant; and finally, 
the lattice formulation of gauge theories, which so far provides
the only convincing theoretical evidence of confinement
and chiral symmetry breaking in non-abelian gauge theories.

\vskip 20pt

\section{Ultraviolet Divergences and Perturbative Non-renormalizability}
\label{sec:one-loop}

In gravity the coupling is dimensionful, $G \sim \mu^{2-d}$, and 
one expects trouble in four dimensions already on purely dimensional grounds,
with divergent one loop corrections proportional to
$G \Lambda^{d-2} $ where $\Lambda$ is the ultraviolet cutoff.
Phrased differently, one expects to lowest order some seriously bad ultraviolet behavior from the running of Newton's constant at large momenta,
\beq
G(k^2) \, / \,  G \,  \sim \, 1 + {\rm const.} \; G \, k^{d-2} + \, O(G^2)
\eeq
While problematic in four dimensions,
these considerations also suggest that ordinary Einstein gravity
should be perturbatively renormalizable in the traditional sense in
two dimensions, an issue to which we will return later.

The more general argument for perturbative non-renormalizability
goes as follows. 
The gravitational action contains the scalar curvature $R$ which
involves two derivatives of the metric.
Thus the graviton propagator in momentum space will go like $1/k^2$,
and the vertex functions like $k^2$.
In $d$ dimensions each loop integral with involve a momentum
integration $d^d k$, so that the superficial degree of divergence ${\cal D}$
of a Feynman diagram with $L$ loops
will be given by
\beq
{\cal D} = 2 + (d-2) \, L
\eeq
independent of the number of external lines.
One concludes that for $d>2$ the degree of ultraviolet 
divergence increases with increasing loop order $L$.

The most convenient tool to determine the structure of the
divergent one-loop corrections to Einstein gravity is the
background field method combined
with dimensional regularization, wherein ultraviolet divergences
appear as poles in $\epsilon=d-4$.
In non-Abelian gauge theories the background field method greatly
simplifies the calculation of renormalization factors, while
at the same time maintaining explicit gauge invariance.
The essence of the method is easy to describe: one replaces
the original field appearing in the classical action by $A+Q$,
where $A$ is a classical background field and $Q$ the quantum fluctuation.
A suitable gauge condition is chosen (the background gauge),
such that manifest gauge invariance is preserved for the
background $A$ field.
After expanding out the action to quadratic order in the $Q$ field,
the functional integration over $Q$ is performed, leading
to an effective action for the background $A$ field.
This method eventually determines, after a rather lengthy calculation, the required one-loop counterterm for pure gravity
\beq
\Delta {\cal L}_g = { \sqrt{g} \over 8 \pi^2 (d-4) }
\left ( {1 \over 120 } R^2 + { 7 \over 20 } R_{\mu\nu} R^{\mu\nu} \right )
\label{eq:one-loop-div}
\eeq

There are two interesting, and interrelated, aspects of the result
of Eq.~(\ref{eq:one-loop-div}).
The first one is that for pure gravity the divergent part vanishes
when one imposes the tree-level equations of motion $R_{\mu\nu}=0$:
the one-loop divergence vanishes on-shell.
The second interesting aspect is that the specific structure of the
one-loop divergence is such that its effect can actually
be re-absorbed into a field redefinition,
\beq
g_{\mu\nu} \; \rightarrow  \; g_{\mu\nu} + \delta g_{\mu\nu} 
\;\;\;\;\;\;  \delta g_{\mu\nu} \, \propto \, 
{ 7 \over 20 } R_{\mu\nu} - { 11 \over 60 } R \, g_{\mu\nu}
\eeq
which renders the one-loop amplitudes finite for pure gravity.
It appears though that these two aspects are largely coincidental;
unfortunately this hoped-for mechanism does not seem to work to two loops,
and no additional miraculous cancellations seem to occur there.

One can therefore attempt to summarize the (perturbative) situation so far as follows: 
In principle perturbation theory in $G$ in provides a clear, covariant 
framework in which radiative corrections to gravity can be computed in a 
systematic loop expansion.
The effects of a possibly non-trivial gravitational measure do not show up
at any order in the weak field expansion, and radiative
corrections affecting the renormalization of the cosmological constant,
proportional to $\delta^d (0)$, are set to zero in dimensional regularization.

At the same time, at every order in the loop expansion new invariant
terms involving higher derivatives of the metric are generated,
whose effects cannot simply be absorbed into a re-definition of
the original couplings.
As expected on the basis of power-counting arguments, the theory
is not perturbatively renormalizable in the traditional sense in four dimensions
(although it seems to fail this test by a small measure in lowest
order perturbation theory).

Thus the standard approach based on a perturbative expansion of the pure Einstein theory in four dimensions is clearly not convergent 
(it is in fact badly divergent), and represents therefore a temporary dead end.
The key question is therefore if this is an artifact of naive perturbation theory,
or not.

\vskip 20pt

\section{Feynman Path Integral for Quantum Gravitation}
\label{sec:path}

If non-perturbative effects play an important role in quantum gravity, 
then one would expect the need for an improved formulation
of the quantum theory is, which would not rely exclusively
on the framework of a perturbative expansion.
After all, the fluctuating metric field $g_{\mu\nu}$ is dimensionless,
and carries therefore no natural scale.
For the somewhat simpler cases of a scalar field and non-Abelian gauge theories 
a consistent
non-perturbative formulation based on the Feynman path integral 
has been known for some time, and is by now well developed.
In a nutshell, the Feynman path integral formulation for pure quantum
gravitation can be expressed in the functional integral formula
\beq
Z = \int_ {\rm geometries }  e^{ \, { i \over \hbar} I_{\rm geometry} } \;\; ,
\label{eq:fey-path}
\eeq
just like the Feynman path integral for a non-relativistic quantum mechanical 
particle expresses
quantum-mechanical amplitudes in terms of sums over paths
\beq
A  ( i \rightarrow f ) = \int_ {\rm paths }  
e^{ { \, i \over \hbar} I_{\rm path} } \;\; .
\eeq
What is the precise meaning of the expression in Eq.~(\ref{eq:fey-path})?
In the case of quantum fields, one is generally interested in
a vacuum-to-vacuum amplitude, which requires
$t_i \rightarrow - \infty $ and $t_f \rightarrow + \infty $.
For a scalar field the functional integral with sources is generally of the form
\beq
Z [J] = \int [d \phi ] \exp \left 
\{ i \int d^4 x [ {\cal L}(x) + J(x) \phi (x) ] \right \}
\label{eq:z-mink}
\eeq
where $[d \phi] = \prod_x d \phi(x) $, and ${\cal L}$ 
the usual Lagrangian density for a scalar field.
It is important to note that even with an underlying lattice discretization,
the integral in Eq.~(\ref{eq:z-mink})
is in general ill-defined without a damping factor, 
due to the overall $i$ in the exponent.
Advances in axiomatic field theory indicate that
if one is able to construct a well defined field theory in Euclidean
space $x=(\bf x, \tau)$ obeying certain axioms, then there
is a corresponding field theory in Minkowski space $({\bf x}, t)$ 
$ t = - \, i \, \tau $ defined as an analytic continuation of the Euclidean theory,
such that it obeys the Wightmann axioms. 

Turning to the case of gravity, it should be clear that at least  to all orders
in the weak field expansion there is really no difference of
substance between the Lorentzian (or pseudo-Riemannian) and the
Euclidean (or Riemannian) formulation.
Indeed most, if not all, of the perturbative calculations of the
preceding section could have been carried out with the 
Riemannian weak field expansion about flat Euclidean space
$ g_{\mu\nu} = \delta_{\mu\nu} + h_{\mu\nu} $
with signature $++++$, or about some suitable classical Riemannian
background manifold. 
Now in function space one needs a metric
before one can define a volume element.
Therefore, following DeWitt, one 
needs first to define an invariant norm for metric deformations  
\beq
\Vert \delta g \Vert^2 \, = \, 
\int d^d x \, \delta g_{\mu \nu}(x) \,
G^{\mu \nu, \alpha \beta} \bigl ( g(x) \bigr ) \,
\delta g_{\alpha \beta}(x) \;\; ,
\label{eq:dw-def}
\eeq
with the supermetric $G$ given by the ultra-local
expression
\beq
G^{\mu \nu, \alpha \beta} \bigl ( g(x) \bigr ) \, = \, 
\half \, \sqrt{g(x)} \, \left [ \,
g^{\mu \alpha}(x) g^{\nu \beta}(x) +
g^{\mu \beta}(x) g^{\nu \alpha}(x) + \lambda \,
g^{\mu \nu}(x) g^{\alpha \beta}(x) \, \right ]
\label{eq:dw-super}
\eeq
with $\lambda$ a real parameter, $\lambda \neq - 2 / d $.
The DeWitt supermetric then defines a suitable volume element $\sqrt{G}$
in function space, such that the functional measure over
the $g_{\mu\nu}$'s taken on the form
\beq
\int [d \, g_{\mu\nu} ] \, \equiv \, \int \, \prod_x \, 
\Bigl [ \, \det G(g(x)) \, \Bigr ]^{1/2} \,
\prod_{\mu \geq \nu} d g_{\mu \nu} (x) \;\; .
\label{eq:dw-det0}
\eeq
Thus the local measure for the Feynman path integral for pure gravity
is given by
\beq
\int \, \prod_x \, \bigl [ g(x) \bigr ]^{ (d-4)(d+1)/8 } \,
\prod_{\mu \ge \nu} \, d g_{\mu \nu} (x)
\label{eq:dw-dewitt}
\eeq
In four dimensions this becomes simply
\beq
\int [d \, g_{\mu\nu} ] \,  = \, 
\int \, \prod_x \, \prod_{\mu \ge \nu} \, d g_{\mu \nu} (x)
\label{eq:dw-dewitt-4d}
\eeq
However it is not obvious that the above construction is unique.
Amore general measure would contain the additional volume factor 
$g^{\sigma /2}$ 
in a slightly more general gravitational functional measure 
\beq
\int [d \, g_{\mu\nu} ] \,  = \, 
\prod_x \, \left [ g(x) \right ]^{\sigma / 2} \, 
\prod_{ \mu \ge \nu } \, d g_{ \mu \nu } (x) \;\; ,
\label{eq:gen-meas}
\eeq
Therefore it is important in this context that one can show that the 
gravitational functional measure
of Eq.~(\ref{eq:gen-meas}) is invariant under infinitesimal general
coordinate transformations, irrespective of the value of $\sigma$.

So in conclusion, the Euclidean Feynman path integral for pure
Einstein gravity with a cosmological constant term is given by 
\beq
Z_{cont} \; = \; \int [ d \, g_{\mu\nu} ] \; \exp \Bigl \{
- \lambda_0 \, \int d x \, \sqrt g \, + \, 
{ 1 \over 16 \pi G } \int d x \sqrt g \, R \Bigr \} \;\; .
\label{eq:zcont}
\eeq
Still not all is well.
Euclidean quantum gravity suffers potentially from a disastrous problem
associated with the conformal instability: the presence of kinetic
contributions to the linearized action entering with the wrong sign.
If one writes down a path integral for pure gravity in the form
of Eqs.~(\ref{eq:zcont})
one realizes that it appears ill defined due to the fact that the scalar
curvature can become arbitrarily positive.
In turn this can be seen as related to the fact that while
gravitational radiation has positive energy, gravitational potential
energy is negative, because gravity is attractive.
To see more clearly that the gravitational action can be made arbitrarily
negative consider the conformal transformation 
$ \tilde g_{\mu \nu} = \Omega ^2 g_{\mu \nu} $ where $\Omega$ 
is some positive function.
Then the Einstein action transforms into
\beq
I_E ( \tilde g ) = - { 1 \over 16 \pi G }  \int  d ^ 4 x  \sqrt g \;
( \Omega^2 R + 6 \; g^{\mu \nu} \partial_\mu \Omega \, \partial_\nu \Omega ) \;\; .
\eeq
which can be made arbitrarily negative by choosing a rapidly varying conformal
factor $\Omega$.
Indeed in the simplest case of a metric $g_{\mu\nu} = \Omega ^2 \eta_{\mu \nu}$
one has 
\beq
\sqrt{g} \, ( R -2 \lambda ) \, = \,  6 \, g^{\mu \nu} \partial_\mu \Omega \, \partial_\nu \Omega  \, - \, 2 \lambda \Omega^4
\eeq
which looks like a $\lambda \phi^4$ theory but with the wrong sign for the
kinetic term.
The problem is referred to as the conformal instability of the classical
Euclidean gravitational action.

A possible solution to the unboundedness problem of the Euclidean theory
is that perhaps it should not be regarded  as necessarily an obstacle to 
defining a quantum theory non-perturbatively.
After all the quantum mechanical attractive Coulomb well problem has, 
for zero orbital
angular momentum or in the one-dimensional case, a similar type of instability,
since the action there is also unbounded from below.
The way the quantum mechanical treatment ultimately evades the problem
is that the particle has a vanishingly small probability amplitude to
fall into the infinitely deep well.
In quantum gravity the question regarding the conformal instability 
can then be rephrased in a similar way: Will the quantum fluctuations 
in the metric be strong enough so that physical excitations will
not fall into the conformal well?
Of course to answer such questions satisfactorily one needs a
formulation which is not restricted to small fluctuations, perturbation theory
and the weak field limit.
Ultimately in the lattice theory the answer seems yes, at least for
sufficiently strong coupling $G$.

\vskip 20pt

\section{Perturbatively Non-renormalizable Theories: The Sigma Model}
\label{sec:sigma}

Einstein gravity is not perturbatively renormalizable in the traditional sense
in four dimensions.
Concretely this means that to one-loop order higher derivative terms are generated as radiative corrections with divergent coefficients.
The natural question then arises: Are there any other
field theories where the standard perurbative treatment fails, 
yet for which one can find alternative methods and from them
develop consistent predictions?
The answer seems unequivocally yes.
Indeed outside of gravity, there are two notable examples of field theories, the non-linear sigma model and the self-coupled fermion model, which are
not perturbatively renormalizable for $d>2$, and yet lead to
consistent, and in some instances testable, predictions above $d=2$.

The key ingredient to all of these results is, as originally recognized by Wilson, the existence of a {\it non-trivial ultraviolet
fixed point} (a phase transition in statistical field theory),
with non-trivial universal scaling dimensions.
Furthermore, one is lucky enough that for the non-linear $\sigma$-model three quite different theoretical approaches are available
for comparing quantitative predictions: the $2+\epsilon$ expansion, the large-$N$
limit, and the lattice approach.
Within the lattice approach, several additional techniques become available:
the strong coupling expansion, the weak coupling expansion, real space
renormalization group methods, and the
numerically exact evaluation of the path integral.
Finally, the results for the non-linear sigma model in the
scaling regime around the non-trivial ultraviolet fixed point can be compared
to recent high accuracy satellite (space shuttle) experiments on three-dimensional systems, 
and the results agree, the
$O(2)$ non-linear $\sigma$-model in three dimensions, 
in some cases to several decimals,
providing one of the most accurate tests of quantum field theory to date!

Concretely, the non-linear $\sigma$-model is a simple model 
describing the dynamics of an $N$-component field $\phi_a$
satisfying a unit constraint $\phi^2(x)=1$,
and with functional integral given by
\beq
Z [J] \, = \, \int [ \, d \phi \, ] \, 
\prod_x \, \delta \left [ \phi (x) \cdot \phi (x) - 1 \right ] \,
\exp \left ( - \, { \Lambda^{d-2} \over g } \, S(\phi) 
\, + \int d^d x \; J(x) \cdot \phi(x) \,  \right )
\label{eq:sigma-cont}
\eeq
The action is taken to be $O(N)$-invariant
\beq
S(\phi) \, = \, \half \,
\int d^d x \; \partial_\mu \phi (x) \cdot \partial_\mu \phi (x)
\eeq
$\Lambda$ here is the ultraviolet cutoff and $g$ the bare
dimensionless coupling at the cutoff scale $\Lambda$;
in a statistical field theory context $g$ plays the role of a temperature,
and $\Lambda$ is proportional to the inverse lattice spacing.
Above two dimensions, $d-2=\epsilon >0$ and a  
perturbative calculation determines the coupling renormalization.
One finds for the effective coupling $g_e$ using dimensional regularization
\beq
{ 1 \over g_e } \; = \;  { \Lambda^\epsilon \over g }
\, \left [ 1 \, - \, { 1 \over \epsilon } \, { N - 2 \over 2 \pi } \, g  
\, + \, O(g^2) \, \right ]
\label{eq:g-nonlin}
\eeq
This then gives immediately the Callan-Symanzik $\beta$-function 
for $g$
\beq
\Lambda \; { \partial \, g \over \partial \, \Lambda } \; = \; 
\beta (g) \; = \; 
\epsilon \, g - { N - 2 \over 2 \pi } \, g^2 
\, + \, O \left ( g^3, \epsilon g^2 \right ) 
\label{eq:beta-nonlin}
\eeq
which determines the scale dependence
of $g ( \mu ) $ for an arbitrary momentum scale $\mu$.
The scale dependence of $g(\mu)$ is such that if the initial
$g$ is less than the ultraviolet fixed point value $g_c$, with
\beq
g_c \, = \, { 2 \pi \epsilon \over N - 2 } \, + \, \dots
\label{eq:gc}
\eeq
then the coupling will flow towards the Gaussian fixed
point at $g=0$.
The new phase that appears when $\epsilon >0$ and
corresponds to a low temperature, spontaneously
broken phase with finite order parameter.
On the other hand if $ g > g_c$ then the coupling
$g(\mu)$ flows towards
increasingly strong coupling, and eventually
out of reach of perturbation theory.
In two dimensions the $\beta$-function has no zero
and only the strong coupling phase is
present.

The one-loop running of $g$ as a function of a sliding momentum scale $\mu=k$
and $\epsilon>0$ can be obtained by integrating 
Eq.~(\ref{eq:beta-nonlin}),
\beq
g(k^2) \; = \; { g_c \over 1 \, \pm \, a_0 \, (m^2 / k^2 )^{(d-2)/2} } 
\label{eq:grun-nonlin} 
\eeq
with $a_0$ a positive constant and $m$ a mass scale;
the combination $ a_0 \, m^{d-2}$ is just the integration
constant for the differential equation. 
The choice of $+$ or $-$ sign is determined from whether one is
to the left (+), or to right (-) of $g_c$, in which case
$g(k^2)$ decreases or, respectively, increases as one flows away
from the ultraviolet fixed point.
It is crucial to realize that
the renormalization group invariant mass scale $\sim m$
arises here as an arbitrary
integration constant of the renormalization group equations,
and cannot be determined from perturbative arguments alone.
One can integrate the $\beta$-function equation 
in Eq.~(\ref{eq:beta-nonlin}) to obtain the renormalization
group invariant quantity
\beq
\xi^{-1} (g) = m(g) = {\rm const.} \; \Lambda \, 
\exp \left ( - \int^g { dg' \over \beta (g') } \right )
\label{eq:xi-beta}
\eeq  
which is identified with the correlation length appearing in
physical correlation functions.
The multiplicative constant in front of the expression on the right
hand side arises as an integration constant, and
cannot be determined from perturbation theory in $g$. 
In the vicinity of the fixed point at $g_c$ one can do the
integral in Eq.~(\ref{eq:xi-beta}), using Eq.~(\ref{eq:gc})
and the resulting linearized expression for the
$\beta$-function in the vicinity of the non-trivial
ultraviolet fixed point,
\beq
\beta (g) 
\; \mathrel{\mathop\sim_{ g \rightarrow g_c }} \;
\beta ' (g_c) \, (g - g_c) \, + \, \dots 
\label{eq:beta-lin}
\eeq
and one finds
\beq
\xi^{-1} (g) = m(g) \propto \,
\Lambda \,  | \, g - g_c \, |^{\nu }
\label{eq:m-sigma}
\eeq
with a correlation length exponent 
$\nu = - 1 / \beta'(g_c) \sim 1 / (d-2) + \dots$.
Thus the correlation length $\xi(g)$
diverges as one approaches the fixed point at $g_c$. 

It is important to note that the above results can be tested experimentally.
A recent sophisticated space shuttle experiment
(Lipa et al 2003) has succeeded in measuring the specific heat
exponent $\alpha=2-3\nu$ of superfluid Helium (which is supposed
to share the same universality class as the $N=2$ 
non-linear $\sigma$-model, with the complex phase of
the superfluid condensate acting as the order parameter)
to very high accuracy 
\beq
\alpha = - 0.0127 (3)
\eeq 
Previous theoretical predictions for the $N=2$ model
include the most recent
four-loop $4-\epsilon$ continuum result $\alpha = - 0.01126 (10)$,
a recent lattice Monte Carlo estimate $\alpha = -0.0146(8)$, 
and the lattice variational renormalization group
prediction $\alpha=-0.0125 (39)$.
One more point that should be mentioned here is
that in the large $N$ limit the non-linear $\sigma$-model can be solved exactly.
This allows an independent verification of the correctness of the 
general ideas presented earlier, as well as a
direct comparison of explicit results for universal quantities.
The general shape of $\beta(g)$ is of
the type shown in Fig. 1.,
with $g_c$ a stable non-trivial UV fixed point, and $g=0$ and $g=\infty$
two stable (trivial) IR fixed points.

Perhaps the core message one gains from the
discussion of the non-linear $\sigma$-model in $d>2$ is that:

The model provides a specific example of a theory which is not
perturbatively renormalizable in the traditional sense, and
for which the naive perturbative expansion in fixed dimension
leads to uncontrollable divergences and inconsistent results;

Yet the model can be constructed perturbatively
in terms of a double expansion in $g$ and $\epsilon=d-2$.
This new perturbative expansion, combined with the renormalization
group, in the end provides explicit and detailed information
about universal scaling properties of the theory in the
vicinity of the non-trivial ultraviolet point at $g_c$;

And finally, that the continuum field theory predictions obtained
this way generally agree, for distances much larger than
the cutoff scale, with lattice results, and,
perhaps most importantly, with high precision experiments on systems 
belonging to the same universality class of the $O(N)$
model. 
Indeed the theory results provide one of the most accurate
predictions of quantum field theory to date!

\begin{center}
\epsfxsize=8cm
\epsfbox{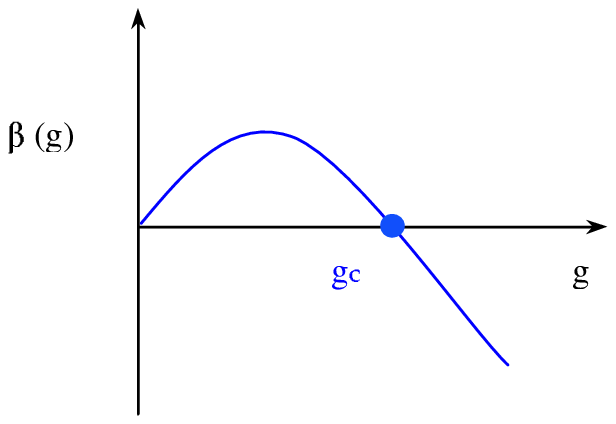}
\end{center}

\noindent{\small Figure 1.
The $\beta$-function for the non-linear $\sigma$-model
in the $2+\epsilon$ expansion and in the large-$N$ limit, for $d>2$.}
\medskip

\label{fig:beta-nonlin}

\vskip 20pt

\section{Phases of Gravity in $2+\epsilon$ Dimensions}
\label{sec:phaseseps}

Can any of these lessons be applied to gravity?
In two dimensions the gravitational coupling becomes dimensionless, 
$G\sim \Lambda^{2-d}$,
and the theory appears therefore perturbatively renormalizable.
In spite of the fact that the gravitational action reduces to a topological
invariant in two dimensions, it would seem meaningful to try to construct, 
in analogy to what was suggested originally by Wilson for scalar field theories, 
the theory perturbatively as a double series in $\epsilon=d-2$ and $G$.
One first notices though that in pure Einstein gravity, with Lagrangian density 
\beq
{\cal L} = - { 1 \over 16 \pi G_0} \, \sqrt{g} \, R \;\; ,
\eeq
the bare coupling $G_0$ can be completely reabsorbed by a field redefinition
\beq
g_{\mu\nu} = \omega \, g_{\mu\nu}'
\label{eq:metric-scale}
\eeq
with $\omega$ is a constant, and thus
the renormalization properties of $G_0$ have no physical meaning
for this theory.
The situation changes though when one introduces a second dimensionful
quantity to compare to.
In the pure gravity case this contribution is naturally
supplied by the cosmological constant term proportional to $\lambda_0$,
\beq
{\cal L} = - { 1 \over 16 \pi G_0} \, \sqrt{g} \, R \, + \, \lambda_0 \sqrt{g}
\eeq
Under a rescaling of the metric as in Eq.~(\ref{eq:metric-scale}) one
has
\beq
{\cal L} = - { 1 \over 16 \pi G_0} \, \omega^{d/2-1} \, \sqrt{g'} \, R' 
\, + \, \lambda_0 \, \omega^{d/2} \, \sqrt{g'}
\label{eq:rescale}
\eeq
which is interpreted as a rescaling of the two bare couplings
\beq
G_0 \rightarrow \omega^{-d/2+1} G_0 \; , \;\;\;\; 
\lambda_0 \rightarrow \lambda_0 \, \omega^{d/2}
\eeq
leaving the dimensionless combination $G_0^d \lambda_0^{d-2}$ unchanged.
Therefore only the latter combination has physical meaning in pure gravity.
In particular, one can always choose the scale $\omega = \lambda_0^{-2/d}$
so as to adjust the volume term to have a unit coefficient.
The $2+\epsilon$ expansion for pure gravity then proceeds as follows.
First the gravitational part of the action
\beq
{\cal L} = - { \mu^\epsilon \over 16 \pi G} \, \sqrt{g} \, R \;\; ,
\label{eq:l-pure}
\eeq
with $G$ dimensionless and $\mu$ an arbitrary momentum scale, 
is expanded by setting
\beq
g_{\mu\nu} \, \rightarrow \, \bar g_{\mu\nu} = g_{\mu\nu} \, + \, h_{\mu\nu}
\eeq
where $g_{\mu\nu}$ is the classical background field and $h_{\mu\nu}$
the small quantum fluctuation.
The quantity ${\cal L}$ in Eq.~(\ref{eq:l-pure}) is naturally identified with
the bare Lagrangian, and the scale $\mu$ with a microscopic ultraviolet
cutoff $\Lambda$, the inverse lattice spacing in a lattice formulation.
Since the resulting perturbative expansion is generally reduced to
the evaluation of Gaussian integrals, the original constraint (in the Euclidean
theory)
\beq
\det g_{\mu\nu} (x) \, > 0 
\label{eq:vol-cont}
\eeq
is no longer enforced (the same is {\it not} true in the lattice
regulated theory, where it plays an important role.
In order to perform the perturbative calculation of the one-loop
divergences a gauge fixing term needs to be added, in the form of a 
background harmonic gauge condition,
\beq
{\cal L}_{gf} = \half \alpha 
\sqrt{ g} \,  g_{\nu\rho}
\left ( 
\nabla_\mu h^{\mu\nu} - \half \beta g^{\mu\nu} \nabla_\mu h 
\right )
\left ( 
\nabla_\lambda h^{\lambda\rho} - 
\half \beta g^{\lambda\rho} \nabla_\lambda h 
\right )
\label{eq:l-gauge}
\eeq
with $ h^{\mu\nu} = g^{\mu\alpha} g^{\nu\beta} h_{\alpha\beta} $,
$h = g^{\mu\nu} h_{\mu\nu}$ and $ \nabla_\mu $ the 
covariant derivative with respect to the background metric
$g_{\mu\nu}$; here $\alpha$ and $\beta$ are some gauge
fixing parameters.
The gauge fixing term also gives rise to a Faddeev-Popov ghost
contribution ${\cal L}_{ghost}$ containing the ghost field
$\psi_\mu$, so that the total
Lagrangian becomes ${\cal L}+{\cal L}_{gf}+{\cal L}_{ghost}$.
After the dust settles, the 
one-loop radiative corrections modify the total Lagrangian to
\beq
{\cal L} \rightarrow
- { \mu^\epsilon \over 16 \pi G} \left ( 1 - { b \over \epsilon } G \right )
\sqrt{g} R 
+ \lambda_0 \left [ 1 - \left (  
{ a_1 \over \epsilon } + { a_2 \over \epsilon^2 } \right ) G \right ] \! \sqrt{g}
\eeq
where $a_1$ and $a_2$ are some constants.
Next one can make use of the freedom to rescale the metric,
by setting
\beq
\left [ 1 - \left (  
{ a_1 \over \epsilon } + { a_2 \over \epsilon^2 } \right ) G \right ] \sqrt{g}
= \sqrt{g'}
\label{eq:scaled-cosm}
\eeq
which restores the original unit coefficient for the cosmological constant term.
The rescaling is achieved by a suitable field redefinition
\beq
g_{\mu\nu} = 
\left [ 1 - \left (  
{ a_1 \over \epsilon } + { a_2 \over \epsilon^2 } \right ) G \right ]^{-2/d} 
\, g'_{\mu\nu}
\eeq
Hence the cosmological term is brought back into the standard form
$\lambda_0 \sqrt{g'}$, and one obtains for the
complete Lagrangian to first order in $G$
\beq
{\cal L} \rightarrow
- { \mu^\epsilon \over 16 \pi G} 
\left [ 1 - { 1 \over \epsilon } ( b - \half a_2 ) G \right ]
\sqrt{g'} R' 
+ \lambda_0 \sqrt{g'}
\eeq
where only terms singular in $\epsilon$ have been retained.
In particular one notices that only the combination $ b - \half a_2 $
has physical meaning, and can in fact be shown to be gauge 
independent.
From this last result one can finally read off the renormalization of 
Newton's constant
\beq
{ 1 \over G} \rightarrow { 1 \over G }
\left [ 1 - { 1 \over \epsilon } ( b - \half a_2 ) \, G \right ]
\label{eq:g-ren}
\eeq
The $a_2$ contribution cancels out the gauge-dependent
part of $b$, giving for the remaining contribution
$ b - \half a_2 = \twoth \cdot 19 $.

In the presence of an explicit renormalization scale parameter
$\mu$ the $\beta$-function for pure gravity is obtained by
requiring the independence of the quantity $G_e$ (here
identified as an effective coupling constant, with lowest order
radiative corrections included) from the original
renormalization scale $\mu$,
\bea
\mu \, { d \over d \mu } \, G_e & = & 0
\nonumber \\
{ 1 \over G_e} & \equiv & { \mu^\epsilon \over G (\mu) }
\left [ 1 - { 1 \over \epsilon } ( b - \half a_2 ) \, G (\mu) \right ]
\label{eq:g-oneloop}
\eea
To first order in $G$, one has from Eq.~(\ref{eq:g-oneloop}) 
\beq
\mu { \partial \over \partial \mu } \, G =
\beta (G) = \epsilon \, G \, - \, \beta_0 \, G^2 \, + \,
O( G^3, G^2 \epsilon )
\label{eq:beta-oneloop}
\eeq 
with, by explicit calculation, $\beta_0 = \twoth \cdot 19 $.
From the procedure outlined above it is clear that $G$ is
the only coupling that is scale-dependent in pure gravity.

Matter fields can be included as well.
When $N_S$ scalar fields and $N_F$ Majorana fermion fields
are added, the results of Eqs.~(\ref{eq:g-ren})
and (\ref{eq:g-oneloop}) are modified to
\beq
b \rightarrow b - \twoth c
\eeq
with $c=N_S + \half N_F$, and therefore for the combined
$\beta$-function of Eq.~(\ref{eq:beta-oneloop}) to one-loop
order one has $\beta_0 = \twoth (19 - c) $.
One noteworthy aspect of the perturbative calculation is
the appearance of a non-trivial ultraviolet fixed point at 
$G_c=(d-2)/\beta_0$ for which $\beta (G_c)=0$,
whose physical significance is discussed below.

In the meantime the calculations have been laboriously extended 
to two loops, with the result
\beq
\mu { \partial \over \partial \mu } G  =
\beta (G) = \epsilon \, G \, - \, \beta_0 \, G^2 
\, - \, \beta_1 \, G^3 \, + \, O( G^4, G^3 \epsilon , G^2 \epsilon^2 )
\label{eq:beta-twoloop}
\eeq 
with $\beta_0 = \twoth \, (25 - c) $ and
$ \beta_1 = { 20 \over 3 } \, (25 - c) $.
The gravitational $\beta$-function of Eqs.~(\ref{eq:beta-oneloop})
and (\ref{eq:beta-twoloop}) determines the scale dependence
of Newton's constant $G$ for $d$ close to two.
It has the general shape shown in Fig. 2.
Because one is left, for the reasons described above, with a
single coupling constant in the pure gravity case, the discussion
becomes in fact quite similar to the non-linear $\sigma$-model case.

\begin{center}
\epsfxsize=8cm
\epsfbox{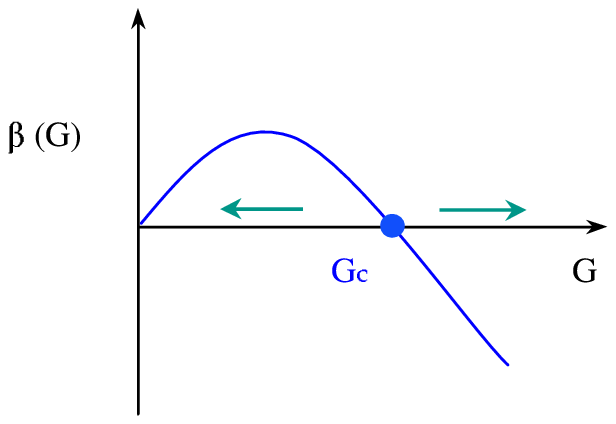}
\end{center}

\noindent{\small Figure 2.
The renormalization group $\beta$-function 
for gravity in $2+\epsilon$ dimensions. 
The arrows indicate the coupling constant flow as one
approaches increasingly larger distance scales.}
\medskip

\label{fig:beta-g-eps}

For a qualitative discussion of the physics it will be simpler in the
following to just focus on the one loop result of Eq.~(\ref{eq:beta-oneloop}); 
the inclusion of the two-loop correction does not alter the qualitative
conclusions by much, as it has the same sign as the lower order, one-loop term.
Depending on whether one is on the right ($G>G_c$) or on the left 
($G<G_c$) of the non-trivial ultraviolet fixed point at
\beq
G_c = { d - 2 \over \beta_0 } + O((d-2)^2 )
\eeq
(with $G_c$ positive provided one has $c<25$)
the coupling will either flow to increasingly larger values of $G$,
or flow towards the Gaussian fixed point at $G=0$, respectively.
The running of $G$ as a function of a sliding momentum scale $\mu=k$
in pure gravity can be obtained by integrating 
Eq.~(\ref{eq:beta-oneloop}), and one has
\beq
G(k^2) \; \simeq \; G_c \, \left [ 
\, 1 \, + \, a_0 \, \left ( {m^2 \over k^2 } \right )^{(d-2)/2}
\, + \, \dots \right ]
\label{eq:grun-cont1} 
\eeq
with $a_0$ a positive constant and $m$ a mass scale.
The choice of $+$ or $-$ sign is determined from whether one is
to the left (+), or to right (-) of $G_c$, in which case
the effective $G(k^2)$ decreases or, respectively, increases as one flows away
from the ultraviolet fixed point towards lower momenta, or larger distances.
Physically the two solutions represent a screening ($G<G_c$) and an 
anti-screening ($G>G_c$) situation.
The renormalization group invariant mass scale $\sim m$
arises here as an arbitrary
integration constant of the renormalization group equations.
At energies sufficiently high to become comparable
to the ultraviolet cutoff, the gravitational coupling $G$ flows towards the
ultraviolet fixed point 
$ G(k^2) \, \mathrel{\mathop\sim_{ k^2 \rightarrow \Lambda^2 }} \, G (\Lambda) $
where $G(\Lambda)$ is the coupling at the cutoff scale $\Lambda$,
to be identified with the bare or lattice coupling. 
Note that the quantum correction
involves a new physical, renormalization group invariant scale $\xi=1/m$
which cannot be fixed perturbatively, and whose size determines the 
scale for the quantum effects.
In terms of the bare coupling $G(\Lambda)$, it is given by
\beq
m \, = \, A_m \cdot \Lambda \, 
\exp \left ( { - \int^{G(\Lambda)} \, {d G' \over \beta (G') } }
\right )
\label{eq:m-cont}
\eeq
which just follows from integrating 
$ \mu { \partial \over \partial \mu }G = \beta (G)$
and then setting $\mu \rightarrow \Lambda$.
The constant $A_m$ on the r.h.s. of Eq.~(\ref{eq:m-cont})
cannot be determined perturbatively, and 
needs to be computed by non-perturbative (lattice) methods.

At the fixed point $G=G_c$ the theory is scale invariant by definition.
In statistical field theory language the fixed point corresponds to a phase transition, where the correlation length $\xi=1/m$ diverges 
and the theory becomes scale (conformally) invariant.
In general in the vicinity of the fixed point, for which $\beta(G)=0$,
one can write
\beq
\beta (G) \, \mathrel{\mathop\sim_{ G \rightarrow G_c }} \, 
\beta' (G_c) \, (G-G_c) \, + \, O ((G-G_c)^2 )
\label{eq:beta-lin-g}
\eeq
If one then defines the exponent $\nu$ by
\beq
\beta ' (G_c) \, = \, - 1/ \nu
\label{eq:nu-def-beta}
\eeq
then from Eq.~(\ref{eq:m-cont}) one has by integration in the vicinity
of the fixed point
\beq
m \, \mathrel{\mathop\sim_{G \rightarrow G_c }} \,
\Lambda \cdot A_m \, | \, G (\Lambda) - G_c |^{\nu} \;\;\; .
\label{eq:m-cont1}
\eeq
which is why $\nu$ is often referred to as the mass gap exponent.
Solving the above equation (with $\Lambda \rightarrow k$)
for $G(k)$ one obtains back Eq.~(\ref{eq:grun-cont1}).
The discussion given above is not altered significantly, at least
in its qualitative aspects, by the inclusion of the
two-loop correction of Eq.~(\ref{eq:beta-twoloop}).
One finds
\bea
G_c & = & {3 \over 2 \, ( 25 - c ) } \, \epsilon
\, - \, {45 \over 2 ( 25 - c )^2 } \, \epsilon^2 \, + \, \dots
\nonumber \\
\nu^{-1} & = & 
\epsilon \, + \, {15 \over 25 - c } \, \epsilon^2 \, + \, \dots
\label{eq:nueps}
\eea
which gives, for pure gravity without matter ($c=0$) in four dimensions, to lowest order
$\nu^{-1} = 2$, and $\nu^{-1} \approx 4.4 $ at the next order.
The key question raised by the perturbative calculations is
therefore: what remains of the above phase transition in four dimensions,
how are the two phases of gravity characterized there non-perturbatively, 
and what is the value of the exponent $\nu$ determining
the running of $G$ in the vicinity of the fixed point 
{\it in four dimensions}.

\vskip 20pt

\section{Lattice Regularized Quantum Gravity}
\label{sec:lattice}

The following section is based on the lattice discretized description
of gravity originally due to Regge, where the Einstein theory is 
expressed in terms of a simplicial decomposition of space-time manifolds.
Its use in quantum gravity is prompted by the desire to make use of techniques
developed in lattice gauge theories,
but with a lattice  which reflects the
structure of space-time rather than just providing a flat passive background.
It also allows one to use powerful nonperturbative analytical techniques
of statistical mechanics as well as numerical methods.

On the lattice the infinite number of degrees of freedom in the continuum
is restricted, by considering Riemannian spaces described
by only a finite number of variables, the geodesic distances between
neighboring points.
Such spaces are taken to be flat almost everywhere and referred to as 
piecewise linear.
The elementary building blocks for $d$-dimensional space-time are
simplices of dimension $d$.
A 0-simplex is a point, a 1-simplex is an edge, a 2-simplex is a triangle, a
3-simplex is a tetrahedron.
A $d$-simplex is a $d$-dimensional object with $d+1$ vertices and
$d(d+1)/2$ edges connecting them.
It has the important property that the values of its edge lengths specify
the shape, and therefore the relative angles, uniquely.  
A simplicial complex can then be viewed as a set of simplices glued together
in such a way that either two simplices are
disjoint or they touch at a common face.
The relative position of points on the lattice is thus completely specified
by an incidence matrix (it tells which point is next to which) and the
edge lengths,
and this in turn induces a metric structure on the piecewise linear space.
Finally the polyhedron constituting the union of all the simplices of
dimension $d$ is called a geometrical complex or skeleton.

Consider a general simplicial lattice in $d$ dimensions, made out
of a collection of flat $d$-simplices glued together at their common faces 
so as to constitute a triangulation of a smooth continuum manifold,
such as the $d$-torus, or the surface of a sphere.
If we focus on one such $d$-simplex, it will itself contain sub-simplices of smaller dimensions; as an example in four dimensions a given 4-simplex will contain 5 tetrahedra, 10 triangles (also referred to as hinges in
four dimensions), 10 edges and 5 vertices.
In general, an $n$-simplex will contain $ { n+1 \choose k+1 } $ $k$-simplices in its boundary.
It will be natural in the following to label simplices by the letter
$s$, faces by $f$ and hinges by $h$.
A general connected, oriented simplicial manifold consisting of $N_s$ $d-$simplices will also be characterized by an incidence matrix
$I_{s,s'}$, whose matrix element $I_{s,s'}$ is chosen to be equal
to one if the two simplices labeled by $s$ and $s'$ share a common face,
and zero otherwise. 

The geometry of the interior of a $d$-simplex is assumed to be flat,
and is therefore completely specified by the lengths of its $d(d+1)/2$ edges.
Let $x^\mu (i)$ be the $\mu$-th coordinate of the $i$-th site.
For each pair of neighboring sites $i$ and $j$ the link length
squared is given by the usual expression
\beq
l^2_{ij} \; = \; \eta_{\mu\nu} \, \left [ x(i)-x(j) \right ]^\mu
\, \left [ x(i)-x(j) \right ]^\nu
\eeq
with $\eta_{\mu\nu}$ the flat metric.
It is therefore natural to associate, within a given
simplex $s$, an edge vector $l_{ij}^{\mu} (s) $
with the edge connecting site $i$ to site $j$. 
When focusing on one such $n$-simplex it will be convenient to label
the vertices by $0,1, 2, 3, \dots , n$ and denote the 
square edge lengths by $l _ {01}^2 = l _ {10}^2$, ... , $l _ {0n}^2 $.
The simplex can then be spanned by the set of $n$ vectors 
$e_1$, ... $e_n$ connecting the vertex $0$ to the 
other vertices.
To the remaining edges within the simplex one then assigns
vectors $e_{ij} = e_i-e_j$ with $1 \le i < j \le n$.
One has therefore $n$ independent vectors, but $\half n (n+1)$
invariants given by all the edge lengths squared within $s$. 
In the interior of a given $n-$simplex one can also assign a second, 
orthonormal (Lorentz) frame, which we will denote in the following 
by $\Sigma (s)$.
The expansion coefficients relating this orthonormal frame to the one specified
by the $n$ directed edges of the simplex associated with the vectors
$e_i $ is the lattice analogue of the $n$-bein or tetrad $e_{\mu}^a$.
Within each $n$-simplex one can define a metric
\beq
g_{ij} (s) \; = \; e_i \cdot e_j \;\; , 
\eeq
with $1 \leq i,j \leq n $, and which in the Euclidean case is positive definite.
In components one has $g_{ij} = \eta_{ab} e_i^a e_j^b$.
In terms of the edge lengths
$l_{ij} \, = \, | e_i - e_ j | $, the metric is given by
\beq
g_{ij} (s) \; = \; \half \, 
\left ( l_{0i}^2 + l_{0j}^2 - l_{ij}^2 \right ) \;\; .
\label{eq:latmet}
\eeq
Comparison with the standard expression for the invariant interval
$ds^2 = g_{\mu\nu} dx^\mu dx^\nu$ confirms that for the metric
in Eq.~(\ref{eq:latmet}) coordinates have been chosen along the
$n$ $e_i$ directions.

The volume of a general $n$-simplex is given by the $n$-dimensional
generalization of the well-known formula for a tetrahedron, namely
\beq
V_n (s) \; = \; {1 \over n ! }  \sqrt { \det  g_{ij} (s) } \;\; .
\label{eq:vol-met}
\eeq
It is possible to associate $p$-forms with lower dimensional
objects within a simplex, which will become useful later.
With each face $f$ of an $n$-simplex (in the shape of a tetrahedron
in four dimensions) one can associate
a vector perpendicular to the face
\beq
\omega (f)_\alpha \; = \; \epsilon_{\alpha \beta_1 \dots \beta_{n-1}} \, 
e^{\beta_1}_{(1)} \dots e^{\beta_{n-1}}_{(n-1)}
\label{eq:omega-face}
\eeq
where $ e_{(1)} \dots e_{(n-1)} $ are a set of oriented edges
belonging to the face $f$, and
$ \epsilon_{\alpha_1 \dots \alpha_n}$ is the sign of the permutation
$ ( \alpha_1 \dots \alpha_n )$.

The volume of the face $f$ is then given by
\beq
V_{n-1} (f) \; = \; \left ( \sum_{\alpha=1}^n  
\omega_\alpha^2 (f) \right )^{1/2}
\eeq
Similarly, one can consider a hinge (a triangle in four dimensions)
spanned by edges $ e_{(1)}$,$\dots$, $ e_{(n-2)}$.
One defines the (un-normalized) hinge bivector
\beq
\omega (h)_{\alpha\beta} \; = \; 
\epsilon_{\alpha \beta \gamma_1 \dots \gamma_{n-2}} \, 
e^{\gamma_1}_{(1)} \dots e^{\gamma_{n-2}}_{(n-2)}
\label{eq:omega-hinge}
\eeq
with the area of the hinge then given by
\beq
V_{n-2} (h) \; = \; {1 \over (n-2)! } 
\left ( \sum_{\alpha < \beta }  \omega_{\alpha\beta}^2 (h) \right )^{1/2}
\eeq
Next, in order to introduce curvature, one needs to define
the dihedral angle between faces in an $n$-simplex.
In an $n$-simplex $s$ two $n-1$-simplices $f$ and $f'$ will intersect
on a common $n-2$-simplex $h$, and the dihedral angle at the specified hinge $h$ is defined as
\beq
\cos \theta (f,f') \; = \; { \omega (f) _{n-1} \cdot \omega (f')_{n-1} \over
V_{n-1}(f) \, V_{n-1} (f') }
\label{eq:dihedralcos}
\eeq
where the scalar product appearing on the r.h.s. can be re-written in terms
of squared edge lengths using
\beq
\omega_{n} \cdot \omega_{n}' \; = \; 
{1 \over (n!)^2 } \, \det ( e_i \cdot e_j' )
\label{eq:omega-dot}
\eeq
and $e_i \cdot e_j'$ in turn expressed in terms of squared edge lengths
by the use of Eq.~(\ref{eq:latmet}).
(Note that the dihedral angle $\theta$ would have to be defined as $\pi$ minus
the arccosine of the expression on the r.h.s. if the orientation
for the $e$'s had been chosen in such a way that the $\omega$'s would
all point from the face $f$ inward into the simplex $s$).  
As an example, in two dimensions and within a given triangle, two edges
will intersect at a vertex, giving $\theta$ as the angle between
the two edges. 
In three dimensions within a given simplex two triangles will intersect
at a given edge, while in four dimension two tetrahedra will
meet at a triangle.
For the special case of an equilateral $n$-simplex, one has simply
$\theta = \arccos {1 \over n}$. 

In a piecewise linear space curvature is detected by going around
elementary loops which are dual to a ($d-2$)-dimensional subspace.
From the dihedral angles associated with the faces of the simplices meeting
at a given hinge $h$ one can compute the deficit angle $\delta (h)$,
defined as
\beq
\delta (h) \; = \; 2 \pi \, - \, \sum_{ s \supset h } \; \theta (s,h)
\label{eq:deficit}
\eeq
where the sum extends over all simplices $s$ meeting on $h$.
It then follows that the deficit angle $\delta$ is a measure
of the curvature at $h$.


Since the interior of each simplex $s$ is assumed to be flat, one can assign
to it a Lorentz frame $\Sigma (s)$.
Furthermore inside $s$ one can define a $d$-component vector 
$\phi (s) = ( \phi_0 \dots \phi_{d-1} )$.
Under a Lorentz transformation of $\Sigma (s)$, described by the
$d \times d$ matrix $\Lambda (s)$ satisfying the usual relation
for Lorentz transformation matrices
\beq
\Lambda^{T} \, \eta \, \Lambda \; = \; \eta
\eeq
the vector $\phi (s)$ will rotate to 
\beq
\phi ' (s) \; = \; \Lambda (s) \, \phi (s)
\eeq
The base edge vectors $e_i^{\mu} = l_{0i}^{\mu} (s)$ themselves
are of course an example of such a vector. 
Next consider two $d$-simplices, individually labeled by $s$ and $s'$,
sharing a common face $f (s,s')$ of dimensionality $d-1$.
It will be convenient to label the $d$ edges residing in the common face
$f$ by indices $i,j=1 \dots d$.
Within the first simplex $s$ one can then assign a Lorentz frame $ \Sigma (s)$,
and similarly within the second $s'$ one can assign the frame $\Sigma (s')$.
The $\half d (d-1) $ edge vectors on the common interface
$f(s,s')$ (corresponding physically to the same edges, viewed from
two different coordinate systems)
are expected to be related to each other by a Lorentz rotation $\bf R$,
\beq
l_{ij}^\mu (s') \; = \; R_{\;\;\nu}^{\mu} (s',s) \; l_{ij}^{\nu} (s) 
\eeq
Under individual Lorentz rotations in $s$ and $s'$ one has of course a 
corresponding change in $\bf R$, namely 
${\bf R} \rightarrow \Lambda (s') \, {\bf R} (s',s) \, \Lambda (s)$.
In the Euclidean $d$-dimensional case $\bf R$ is an orthogonal matrix,
element of the group $SO(d)$.
In the absence of torsion, one can use the matrix ${\bf R}(s',s)$ to describes
the parallel transport of any vector $\phi^\mu$ from simplex $s$ to a
neighboring simplex $s'$,
\beq
\phi^\mu (s') \; = \; R_{\; \; \nu}^{\mu} (s',s) \, \phi^{\nu} (s) 
\eeq
${\bf R}$ therefore describes a lattice version of the connection.
Indeed in the continuum such a rotation would be described by the matrix
\beq
R_{\;\;\nu}^{\mu} \; = \; \left ( e^{\Gamma \cdot dx} \right )_{\;\;\nu}^{\mu}
\eeq
with $\Gamma^{\lambda}_{\mu\nu}$ the affine connection.
The coordinate increment $dx$ is interpreted as joining
the center of $s$ to the center of $s'$, thereby intersecting
the face $f(s,s')$. 
On the other hand, in terms of the Lorentz frames 
$\Sigma (s)$ and $\Sigma (s')$ defined within the two
adjacent simplices, the rotation matrix is given instead by 
\beq
R^a_{\;\;b} (s',s) \; = \; e^a_{\;\;\mu} (s') \, e^{\nu}_{\;\;b} (s)
\; R_{\;\;\nu}^{\mu} (s',s)
\eeq
(this last matrix reduces to the identity if the two orthonormal bases
$\Sigma (s)$ and $\Sigma (s')$ are chosen to be the same,
in which case the connection is simply given by 
$ R(s',s)_{\mu}^{\;\; \nu} = e_{\mu}^{\;\;a} \, e^{\nu}_{\;\;a} $).
Note that it is possible to choose coordinates so that
$ {\bf R} (s,s')$ is
the unit matrix for one pair of simplices, but it will not then be unity for
all other pairs if curvature is present.

This last set of results will be useful later when discussing lattice Fermions.
Let us consider here briefly the problem of how to introduce lattice
spin rotations.
Given in $d$ dimensions the above rotation matrix $ {\bf R} (s',s) $, 
the spin connection ${\bf S}(s,s')$ between two neighboring simplices 
$s$ and $s'$ is defined as follows.
Consider $\bf S$ to be an element of the $2^\nu$-dimensional representation
of the covering group of $SO(d)$, $Spin(d)$, with $d=2 \nu$ or $d=2 \nu+1$, and
for which $S$ is a matrix of dimension $2^\nu \times 2^\nu$.
Then $\bf R$ can be written in general as
\beq
{\bf R} \; = \; \exp \left [ \, 
\half \, \sigma^{\alpha\beta} \theta_{\alpha\beta} \right ]
\eeq
where $\theta_{\alpha\beta}$ is an antisymmetric matrix
The $\sigma$'s are $\half d(d-1)$ $d \times d$ matrices, 
generators of the Lorentz group 
($SO(d)$ in the Euclidean case, and $SO(d-1,1)$ in the Lorentzian case),
whose explicit form is 
\beq
\left [ \sigma_{\alpha\beta} \right ]^{\gamma}_{\;\; \delta}
\; = \; \delta_{\;\; \alpha}^{\gamma} \, \eta_{\beta\delta} \, - \, 
\delta_{\;\; \beta}^{\gamma} \, \eta_{\alpha\delta} \;\; .
\eeq
For fermions the corresponding spin rotation matrix is then obtained from
\beq
{\bf S} \; = \; \exp \left [ \, {\textstyle {i\over4} \displaystyle} \,
\gamma^{\alpha\beta} \theta_{\alpha\beta} \right ]
\eeq
with generators
$ \gamma^{\alpha\beta} = { 1 \over 2 i } [ \gamma^\alpha , \gamma^\beta ] $,
and with the Dirac matrices $\gamma^\alpha$ satisfying as usual  
$\gamma^\alpha \gamma^\beta + \gamma^\beta \gamma^\alpha = 2 \,\eta^{\alpha\beta}$.
Taking appropriate traces, one can obtain a direct relationship
between the original rotation matrix ${\bf R} (s,s')$ and the corresponding
spin rotation matrix ${\bf S}(s,s')$
\beq
R_{\alpha\beta} \; = \; \tr \left ( 
{\bf S}^\dagger \, \gamma_\alpha \, {\bf S} \, \gamma_\beta \right )
/ \tr {\bf 1 }
\label{eq:spinrot}
\eeq
which determines the spin rotation matrix up to a sign.

One can consider a sequence of rotations along an arbitrary
path $P (s_1, \dots , s_{n+1})$ going through simplices 
$s_1 \dots s_{n+1}$, whose combined rotation matrix is given by
\beq
{\bf R} (P) \; = \; {\bf R} (s_{n+1}, s_n ) \cdots {\bf R} (s_2, s_1 )
\eeq
and which describes the parallel transport of an arbitrary vector
from the interior of simplex $s_1$ to the interior of simplex $s_{n+1}$,
\beq
\phi^\mu (s_{n+1}) \; = \; R_{\; \; \nu}^{\mu} (P) \, \phi^{\nu} (s_1) 
\eeq
If the initial and final simplices $s_{n+1}$ and $s_1$ coincide,
one obtains a closed path $C (s_1, \dots , s_n)$, for which the
associated expectation value can be considered as the gravitational
analog of the Wilson loop.
Its combined rotation is given by
\beq
{\bf R} (C) \; = \; {\bf R} (s_1, s_n ) \cdots {\bf R} (s_2, s_1 )
\label{eq:loop-rot}
\eeq
Under Lorentz transformations within each simplex $s_i$ along
the path one has a pairwise cancellation of the $\Lambda (s_i)$ matrices
except at the endpoints, giving in the closed loop case
\beq
{\bf R} (C) \; \rightarrow \; \Lambda ( s_1 ) \, {\bf R} ( C )
\, \Lambda^{T} ( s_1 )
\eeq
Clearly the deviation of the matrix ${\bf R} (C)$ from unity is a measure of curvature.
Also, the trace $\tr {\bf R} (C)$ is independent of the choice
of Lorentz frames.

\begin{center}
\epsfxsize=8cm
\epsfbox{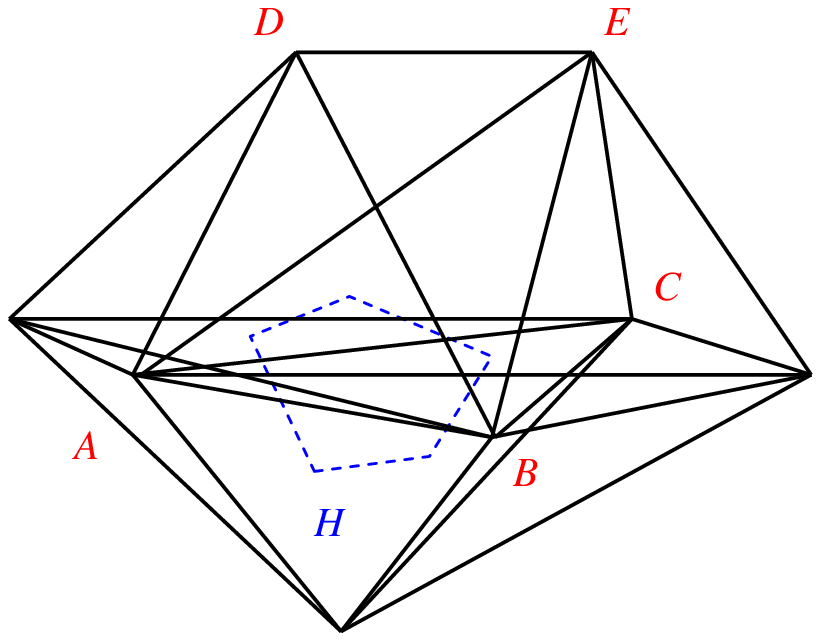}
\end{center}

\noindent{\small Figure 3.
Elementary polygonal path around a hinge
(triangle) in four dimensions. 
The hinge $ABC$, contained in the simplex
$ABCDE$, is encircled by the polygonal path $H$ connecting the
surrounding vertices, which reside in the dual lattice.
One such vertex is contained within the simplex $ABCDE$.}
\medskip

\label{fig:hinge-path}

Of particular interest is the elementary loop associated with
the smallest non-trivial, segmented parallel transport path one can
build on the lattice.
One such polygonal path in four dimensions is shown in 
Fig. 3.
In general consider a $(d-2)$-dimensional simplex (hinge) $h$, which
will be shared by a certain number $m$ of $d$-simplices, 
sequentially labeled by $s_1 \dots s_m$, and whose common faces 
$f(s_1,s_2) \dots f( s_{m-1}, s_m ) $ will also contain the hinge $h$.
Thus in four dimensions several four-simplices will contain,
and therefore encircle, a given triangle (hinge).
In three dimensions the path will encircle an edge, while in two
dimensions it will encircle a site. 
Thus for each hinge $h$ there is a unique elementary closed path $C_h$
for which one again can define the ordered product
\beq
{\bf R} (C_h) \; = \; {\bf R} (s_1, s_m ) \cdots {\bf R} (s_2, s_1 )
\label{eq:loop-rot1}
\eeq
The hinge $h$, being geometrically an object of dimension $(d-2)$, is naturally
represented by a tensor of rank $(d-2)$, referred to a coordinate system
in $h$: an edge vector $l_h^\mu$ in $d=3$, and an area bi-vector 
$\half ( l_h^\mu l_h^{'\nu} - l_h^\nu l_h^{'\mu} ) $ in $d=4$ etc.
Following Eq.~(\ref{eq:omega-hinge}) it will therefore be convenient
to define a hinge bi-vector $U$ in any dimension as
\beq
U_{\mu\nu} (h) \; = \; {\cal N} \, \epsilon_{\mu\nu \alpha_1 \alpha_{d-2}} \, l_{(1)}^{\alpha_1} \dots l_{(d-2)}^{\alpha_{d-2}} \;\; ,
\label{eq:bivector-d}
\eeq
normalized, by the choice of the constant ${\cal N}$, in such a way that 
$U_{\mu\nu} U^{\mu\nu} =2$. 
In four dimensions
\beq
U_{\mu\nu} (h) \; = \; { 1 \over 2 A_h } \;
\epsilon_{\mu\nu\alpha\beta} \, l_1^{\alpha} \, l_2^{\beta}
\label{eq:bivector}
\eeq
where $l_1 (h)$ and $l_2 (h)$ two independent edge vectors
associated with the hinge $h$, 
and $A_h$ the area of the hinge.

An important aspect related to the rotation of an 
arbitrary vector, when parallel transported around a hinge $h$,
is the fact that, due to the hinge's intrinsic orientation,
only components of the vector in the plane perpendicular
to the hinge are affected.
Since the direction of the hinge $h$ is specified locally by
the bivector $U_{\mu\nu}$ of Eq.~(\ref{eq:bivector}), 
one can write for the loop rotation matrix $\bf R$
\beq
R_{\;\;\nu}^{\mu} (C) \; = \; 
\left ( e^{ \delta \, U } \right )_{\;\;\nu}^{\mu}
\label{eq:rot-hinge}
\eeq
where $C$ is now the small polygonal loop entangling the hinge $h$,
and $\delta$ the deficit angle at $h$, previously
defined in Eq.~(\ref{eq:deficit}).
One particularly noteworthy aspect of this last result is the fact that the area
of the loop $C$ does not enter in the expression for the
rotation matrix, only the deficit angle and the hinge direction.
 
At the  same time, in the continuum a vector $V$ carried around
an infinitesimal loop of area $A_C$ will change by 
\beq
\Delta V^{\mu} \; = \; \, \half \, R^{\mu}_{ \;\; \nu \lambda \sigma }
\, A^{ \lambda \sigma } \, V^\nu
\eeq
where $A^{ \lambda \sigma }$ is an area bivector in the plane of $C$,
with squared magnitude $ A_{ \lambda \sigma } A^{ \lambda \sigma } = 2 A_C^2$.
Since the change in the vector $V$ is given by
$ \delta V^\alpha = ({\bf{R-1}})^{\alpha}_{\;\;\beta} \, V^\beta $
one is led to the identification
\beq
\half \; R^{\alpha}_{\;\;\beta\mu\nu} \, A^{\mu\nu} \; = \;
({\bf {R-1}})^{\alpha}_{\;\;\beta} \;\; .
\label{eq:riemrot}
\eeq
Thus the above change in $V$ can equivalently be re-written in terms of the
infinitesimal rotation matrix
\beq
R_{\;\;\nu}^{\mu} (C) \; = \; \left ( e^{ \, \half \, R \cdot A } \right )_{\;\;\nu}^{\mu}
\label{eq:rot-cont}
\eeq
where the Riemann tensor appearing in the exponent on the r.h.s. should not
be confused with the rotation matrix $\bf R$ on the l.h.s..

It is then immediate to see that the two expressions for the rotation
matrix $\bf R$ in Eqs.~(\ref{eq:rot-hinge}) and (\ref{eq:rot-cont})
will be compatible provided one uses for the Riemann tensor
at a hinge $h$ the expression 
\beq
R_{\mu\nu\lambda\sigma} (h) \; = \; {\delta (h) \over A_C (h) } 
\, U_{\mu\nu} (h) \, U_{\lambda\sigma} (h)
\label{eq:riem-hinge}
\eeq
expected to be valid in the limit of small curvatures,
with $A_C (h) $ the area of the loop entangling the hinge $h$.
Here use has been made of the geometric relationship
$U_{\mu\nu} \, A^{\mu\nu} = 2 A_C$.
Note that the bivector $U$ has been defined to be perpendicular
to the $(d-2)$ edge vectors spanning the hinge $h$, 
and lies therefore in the same plane as the loop $C$.
The area $A_C$ is most suitably defined by introducing the
notion of a dual lattice,
i.e. a lattice constructed by assigning centers to the simplices,
with the polygonal curve $C$ connecting these centers sequentially,
and then assigning an area to the interior of this curve.
One possible way of assigning such centers is by introducing
perpendicular bisectors to the faces of a simplex, and locate
the vertices of the dual lattice at their common intersection,
a construction originally discussed by Voronoi.


The first step in writing down an invariant lattice action,
analogous to the continuum Einstein-Hilbert action, 
is to find the lattice analogue of the Ricci scalar.
From the expression for the Riemann tensor at a hinge
given in Eq.~(\ref{eq:riem-hinge}) one obtains by contraction
\beq
R (h) \; = \; 2 \, { \delta (h) \over A_C (h) }
\eeq
The continuum expression $\sqrt{g} \, R$ is then obtained
by multiplication with the volume element $V (h) $ associated with
a hinge.
The latter is defined by first joining the vertices of the
polyhedron $C$, whose vertices lie in the dual lattice,
with the vertices of the hinge $h$, and then computing its volume.

By defining the polygonal area $A_C$ as 
$A_C (h) = d \, V (h) / V^{(d-2)} (h) $, where $V^{(d-2)} (h)$
is the volume of the hinge (an area in four dimensions),
one finally obtains for the Euclidean lattice action for pure gravity
\beq
I_{R} (l^2) \; = \; - \; k \, \sum_{\rm hinges \; h}
\, \delta (h) \, V^{(d-2)} (h) \;\; ,
\label{eq:regge-d}
\eeq
with the constant $k=1/(8 \pi G)$. 
One would have obtained the same result for the single-hinge
contribution to the lattice action
if one had contracted the infinitesimal form of the rotation
matrix $R(h)$ in Eq.~(\ref{eq:rot-hinge}) with the hinge bivector 
$\omega_{\alpha\beta}$ of Eq.~(\ref{eq:omega-hinge}) (or equivalently
with the bivector $U_{\alpha\beta}$ of Eq.~(\ref{eq:bivector})
which differs from $\omega_{\alpha\beta}$ by a constant).
The fact that the lattice action only involves the content of
the hinge $ V^{(d-2)} (h)$ (the area of a triangle in four dimensions)
is quite natural in view of the fact that the rotation matrix
at a hinge in Eq.~(\ref{eq:rot-hinge}) only involves the deficit angle, 
and not the polygonal area $A_C (h)$.

Other terms need to be added to the lattice action.
Consider for example a cosmological constant term, which in the continuum theory
takes the form $ \lambda_0 \int d^d x \sqrt g $.
The expression for the cosmological constant term on the lattice involves
the total volume of the simplicial complex.
This may be written as
\beq
V_{{\bf total}} = \sum_{\rm simplices \; s} V_s
\label{eq:totlatvol}
\eeq
or equivalently as
\beq
V_{{\bf total}} = \sum_{\rm hinges \; h} V_h 
\label{eq:totlatvol1}
\eeq
where $V_h$ is the volume associated with each hinge via the construction
of a dual lattice, as described above.
Thus one may regard the local volume element $ \sqrt g \, d^d x $ as
being represented by either $ V_h $ (centered on $h$) or $ V_s$
(centered on $s$).

The Regge and cosmological constant term then lead to the combined action
\beq
I_{\rm latt} (l^2) \; = \; \lambda_0 \sum_{\rm simplices \; s} \, V^{(d)}_s
\, - \, k \sum_{\rm hinges \; h} \,  \delta_h \, V^{(d-2)}_h
\label{eq:latac}
\eeq


Another interesting aspect is the exact local gauge invariance of the 
lattice action.
Consider the two-dimensional flat skeleton shown in Fig.4.
It is clear that one can move around a point on the surface, keeping
all the neighbors fixed, without violating the triangle inequalities and
leave all curvature invariants unchanged.

\begin{center}
\epsfxsize=8cm
\epsfbox{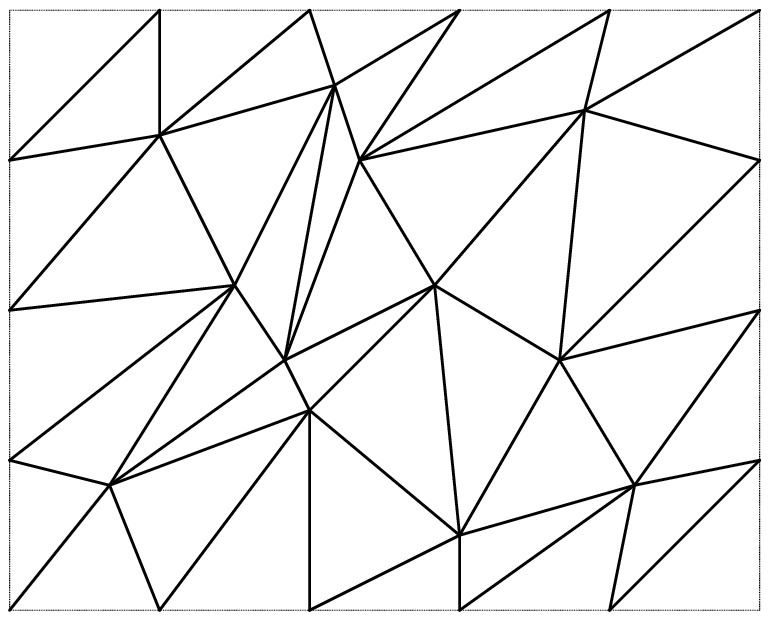}
\end{center}

\noindent{\small Figure 4.
On a random simplicial lattice there are in general
no preferred directions. The lattice can be deformed
locally from one configuration of edges to another which
has the same localized curvature, and illustrates the lattice
analog of the continuum diffeomorphism invariance.}
\medskip

\label{fig:random}

In $d$ dimensions this transformation has $d$ parameters and is an exact
invariance of the action.
When space is slightly curved, the invariance is in general only an approximate
one, even though for piecewise linear spaces piecewise diffeomorphisms
can still be defined as the
set of local motions of points that leave the local contribution to the action,
the measure and the lattice analogues of the continuum curvature invariants unchanged.
Note that in general the gauge deformations of the edges are still constrained
by the triangle inequalities.
The general situation is illustrated in Figure 4.
In the limit when the number of edges becomes very large one expects the full continuum diffeomorphism group to be recovered.
In general the structure of lattice local gauge transformations
is rather complicated and will not be given here. 
These are defined as transformations acting locally on a given
set of edges which leave the local lattice curvature invariant. 
The simplest context in which this local invariance can be exhibited
explicitly is
the lattice weak field expansion.
From the transformation properties of the edge lengths it is clear
that their transformation properties are related to those of
the local metric, as already suggested for example by
the identification of Eqs.~(\ref{eq:latmet}) and (\ref{eq:latmet1}).
In the quantum theory, a local gauge invariance
implies the existence of conservation laws and Ward identities for $n$-point functions.

\section{Lattice Regularized Path Integral}

\label{sec:lattmeas}

As the edge lengths $l_{ij}$ play the role of the continuum metric
$g_{\mu\nu}(x)$, one would expect the discrete measure to involve an
integration over the squared edge lengths.
Indeed the induced metric at a simplex is related to the squared edge
lengths within that simplex, via the expression for the
invariant line element $ds^2 = g_{\mu \nu} dx^\mu dx^\nu$.
After choosing coordinates along the edges emanating from a vertex,
the relation between metric perturbations and squared edge
length variations for a given simplex based at 0 in $d$ dimensions is
\beq
\delta g_{ij} (l^2) \; = \; \half \;
( \delta l_{0i}^2 + \delta l_{0j}^2 - \delta l_{ij}^2 ) \;\; .
\label{eq:latmet1}
\eeq
For one $d$-dimensional simplex labeled by $s$
the integration over the metric is thus equivalent to an 
integration over the edge lengths, and one has the identity
\beq
\left ( {1 \over d ! } \sqrt { \det g_{ij}(s) } \right )^{\sigma} \! \!
\prod_{ i \geq j } \, d g_{i j} (s) = 
{\textstyle \left ( - { 1 \over 2 } \right ) \displaystyle}^{ d(d-1) \over 2 }
\left [ V_d (l^2) \right ]^{\sigma} \!
\prod_{ k = 1 }^{ d(d+1)/2 } \! \! dl_{k}^2
\label{eq:simpmeas}
\eeq
There are $d(d+1)/2$ edges for each
simplex, just as there are $d(d+1)/2$ independent components for the metric
tensor in $d$ dimensions.
Here one is ignoring temporarily the triangle inequality constraints,
which will further require all sub-determinants of $g_{ij}$ to be
positive, including the obvious restriction $l_k^2 >0$.

Let us discuss here briefly the simplicial inequalities which
need to be imposed on the edge lengths.
These are conditions on the edge lengths $l_{ij}$ 
such that the sites $i$ can be considered the vertices of a
$d$-simplex embedded in flat $d$-dimensional Euclidean space.
In one dimension, $d=1$, one requires trivially for all edge lengths
$ l_{ij}^2 > 0 $.
In higher dimensions one requires that
all triangle inequalities and their higher dimensional analogs
to be satisfied, 
\bea
l_{ij}^2 & > & 0
\nonumber \\
V_{k}^2 & = & \left ( {1 \over k ! } \right )^2  \det g_{ij}^{(k)}(s) \, > 0 
\label{eq:tieq-d}
\eea
with $k=2 \dots d$ for every possible choice of sub-simplex
(and therefore sub-determinant) within the original simplex $s$.
The extension of the measure to many simplices glued together
at their common faces is then immediate. 
For this purpose one first needs to identify edges
$ l_k (s) $ and $ l_{k'} (s') $ which are shared between
simplices $s$ and $s'$,
\beq
\int_0^\infty d l^2_k (s) \, \int_0^\infty d l^2_{k'} (s') \;
\delta \left [ l^2_k (s) - l^2_{k'} (s') \right ]
\, = \, \int_0^\infty d l^2_k (s) \;\; .
\eeq
After summing over all simplices one derives,
up to an irrelevant numerical constant, the unique functional measure
for simplicial geometries
\beq
\int [ d l^2] \; = \; 
\int_\epsilon^\infty \; \prod_s \; \left [ V_d (s) \right ]^{\sigma} \;
\prod_{ ij } \, dl_{ij}^2 \; \Theta [l_{ij}^2]  \;\; .
\label{eq:lattmeas}
\eeq
Here $ \Theta [l_{ij}^2] $ is
a (step) function of the edge lengths, with the property
that it is equal to one whenever the triangle inequalities and their
higher dimensional analogs are satisfied,
and zero otherwise.
The quantity $\epsilon$ has been introduced as a cutoff at small
edge lengths.
If the measure is non-singular for small edges, one can safely
take the limit $\epsilon \rightarrow 0$.
In four dimensions the lattice analog of the DeWitt measure
($\sigma=0$) takes on a particularly simple form, namely
\beq
\int [ d l^2] \; = \; \int_0^\infty \prod_{ ij } \, dl_{ij}^2 
\; \Theta [ l_{ij}^2 ] \;\; .
\label{eq:dewlattmeas}
\eeq
Lattice measures over the space of squared edge lengths
have been used extensively in numerical simulations of simplicial
quantum gravity.
The derivation of the above lattice measure
closely parallels the analogous procedure in the
continuum.

The lattice action of Eq.~(\ref{eq:latac}) for pure four-dimensional Euclidean 
gravity then contains
a cosmological constant and Regge scalar curvature term,
as well as possibly higher derivative terms.
It only couples edges which belong either to
the same simplex or to a set of neighboring simplices, and can therefore
be considered as local, just like the continuum action.
It leads to a regularized lattice functional integral
\beq 
Z_{latt} \; = \;  \int [ d \, l^2 ] \; e^{ 
- \lambda_0 \sum_h V_h \, + \, k \sum_h \delta_h A_h } \;\; ,
\label{eq:zlatt} 
\eeq
where, as customary, the lattice ultraviolet cutoff is set equal to one
(i.e. all length scales are measured in units of the lattice cutoff).
The lattice partition function $Z_{latt}$ should then be compared to the
continuum Euclidean Feynman path integral of Eq.~(\ref{eq:zcont}),
\beq
Z_{cont} \; = \; \int [ d \, g_{\mu\nu} ] \; e^{ 
- \lambda_0 \, \int d x \, \sqrt g \, + \, 
{ 1 \over 16 \pi G } \int d x \sqrt g \, R} \;\; ,
\eeq
Occasionally it can be convenient to include the
$\lambda_0$-term in the measure. 
For this purpose one defines 
\beq 
d \mu (l^2) \; \equiv \; [ d \, l^2 ] \; e^{- \lambda_0 \sum_h V_h }
\;\; .
\label{eq:mulatt} 
\eeq
It should be clear that this last expression represents a fairly non-trivial 
quantity, both in view of the relative complexity
of the expression for the volume of a simplex,
and because of the generalized triangle inequality constraints 
already implicit in $[d\,l^2]$.
But, like the continuum functional measure, it is certainly local,
to the extent that each edge length
appears only in the expression for the volume of those simplices
which explicitly contain it.
Furthermore, $\lambda_0$ sets the overall scale and can therefore be set 
equal to one without any loss of generality.

\section{Matter Fields}

\label{sec:scalars}

In the previous section we have discussed the construction
and the invariance properties of a lattice action for pure gravity.
Next a scalar field can be introduced as the simplest type of dynamical
matter that can be coupled invariantly to gravity.
In the continuum the scalar action for a single component field $\phi(x)$
is usually written as 
\beq
I [ g, \phi ] = \half \int d x \, \sqrt g \, [ \,
g^{ \mu \nu } \, \partial_\mu \phi \, \partial_\nu \phi
+ ( m^2 + \xi R ) \phi^2 ] + \dots
\label{eq:scalar}
\eeq
where the dots denote scalar self-interaction terms.
One way to proceed is to introduce a lattice scalar $\phi_i$
defined at the vertices of the simplices.
The corresponding lattice action can then be obtained
through a procedure by which the original continuum metric
is replaced by the induced lattice metric, with the latter
written in terms of squared edge lengths as in Eq.~(\ref{eq:latmet}).
Thus in two dimensions to construct a lattice action for the scalar field, one performs the replacement
\bea
g_{\mu\nu} (x) & \longrightarrow & g_{ij} (\Delta)
\nonumber \\
\det g_{\mu\nu} (x) & \longrightarrow & \det g_{ij} (\Delta)
\nonumber \\
g^{\mu\nu} (x) & \longrightarrow & g^{ij} (\Delta)
\nonumber \\
\partial_\mu \phi \, \partial_\nu \phi & \longrightarrow & 
\Delta_{i} \phi \, \Delta_{j} \phi
\eea
with the following definitions
\beq
g_{ij} (\Delta) = \left( \begin{array}{cc}
l_{3}^2 & \half ( - l_1^2 + l_2^2 + l_{3}^2 ) \cr
\half ( - l_1^2 + l_2^2 + l_{3}^2 ) & l_2^2 \cr
\end{array} \right) \;\; ,
\eeq
\beq
\det g_{ij} (\Delta) = 
\quarter \left [
2 ( l_1^2 l_2^2 + l_2^2 l_{3}^2 + l_{3}^2 l_1^2 ) -
l_1^4 - l_2^4 - l_{3}^4 \right ] \, \equiv \, 4 A_{\Delta}^2 \;\; ,
\eeq
\beq
g^{ij} (\Delta) = 
{ 1 \over \det g (\Delta) } \, \left( \begin{array}{cc}
l_2^2 & \half (l_1^2 - l_2^2 - l_{3}^2 ) \cr
\half (l_1^2 - l_2^2 - l_{3}^2 ) & l_{3}^2 \cr
\end{array} \right) \;\; .
\eeq
The scalar field derivatives get replaced as usual by finite differences
\beq
\partial_\mu \phi \, \longrightarrow \, ( \Delta_\mu \phi )_i \; = \;
\phi_{ i + \mu } - \phi_ i \;\; .
\eeq
where the index $\mu$ labels the possible directions in which one
can move away from a vertex within a given triangle.
After some suitable re-arrangements one finds for the lattice
action describing a massless scalar field
\beq
I (l^2, \phi) \; = \; \half \sum_{<ij>} A_{ij} \,
\Bigl ( { \phi_i - \phi_j \over l_{ij} } \Bigr )^2 \;\; .
\label{eq:acdual} 
\eeq
Here $A_{ij}$ is the dual (Voronoi) area associated with the edge $ij$,
and the symbol $< \! ij \! >$ denotes a sum over nearest neighbor lattice
vertices.
It is immediate to generalize the action of Eq.~(\ref{eq:acdual}) to
higher dimensions, with the two-dimensional Voronoi volumes
replaced by their higher dimensional analogs, leading to
\beq
I (l^2, \phi) \; = \; \half \sum_{<ij>} V_{ij}^{(d)} \,
\Bigl ( { \phi_i - \phi_j \over l_{ij} } \Bigr )^2 \;\; ,
\label{eq:acdual-d} 
\eeq
Here $V_{ij}^{(d)}$ is the dual (Voronoi) volume associated with the edge $ij$,
and the sum is over all links on the lattice.



Spinor fields $\psi_s$ 
and $\bar \psi_s $ are most naturally placed
at the center of each d-simplex $s$.
As in the continuum, the
construction of a suitable lattice action requires the
introduction of the Lorentz group and its associated
tetrad fields $e_\mu^a (s) $ within each simplex labeled
by $s$.  
Within each simplex one can choose a representation of the
Dirac gamma matrices, denoted here by $ \gamma^\mu (s)$,
such that in the local coordinate basis
\beq
\left \{ \gamma^\mu (s) , \gamma^\nu (s) \right \} \; = \; 2 \, g^{\mu\nu} (s)
\eeq
These in turn are related to the ordinary Dirac gamma
matrices $\gamma^a$, which obey
$ \left \{ \gamma^a , \gamma^b \right \} \; = \; 2 \, \eta^{ab} $,
by $ \gamma^\mu (s) \; = \; e^\mu_a (s) \, \gamma^a $.
so that within each simplex the tetrads $e_\mu^a (s) $
satisfy the usual relation
\beq
e_a^\mu (s) \; e_b^\nu (s) \; \eta^{ab} \; = \; g^{\mu\nu} (s)
\eeq 
In general the tetrads are not fixed uniquely within a simplex,
being invariant under the local Lorentz transformations
discussed earlier.
In the continuum the action for a massless spinor field is
given by
\beq
I \; = \; \int d x \sqrt{g} \; \bar \psi (x) \, 
\gamma^\mu \, D_\mu \, \psi (x)
\label{eq:fermac}
\eeq
where $D_\mu = \partial_\mu + \half \omega_{\mu ab} \sigma^{ab}$
is the spinorial covariant derivative containing the spin connection 
$\omega_{\mu ab}$.
On the lattice one then needs a rotation matrix relating 
the vierbeins $e^\mu_a (s_1)$
and $e^\mu_a (s_2)$ in two neighboring simplices.
The matrix ${\bf R} (s_2, s_1)$ is such that
\beq
e_a^\mu (s_2) \; = \; R^\mu_{\;\;\nu} (s_2,s_1) \; e_a^\nu (s_1)
\eeq
and whose spinorial representation $\bf S$
was given previously in Eq.~(\ref{eq:spinrot}).
The invariant lattice action for a massless spinor then takes the simple form
\beq
I \; = \; \half \sum_{\rm faces \; f(s s')} \, V( f(s,s')) \,
\bar \psi_s \, {\bf S} ( {\bf R} (s,s') ) \, \gamma^\mu (s') \,
n_\mu (s,s') \, \psi_{s'}
\eeq
where the sum extends over all interfaces $f(s,s')$ connecting one
simplex $s$ to a neighboring simplex $s'$.
The above spinorial action can be considered analogous to the lattice
Fermion action proposed originally by Wilson for non-Abelian
gauge theories.


For gauge fields a locally gauge invariant action for 
an $SU(N)$ gauge field coupled to gravity is
\beq
I_{\rm gauge} \; = \; - { 1 \over 4 g^2 } \, \int d^4 x \, 
\sqrt{g} \, g^{\mu\lambda} \, g^{\nu\sigma} \,
F^a_{\mu\nu} \, F^a_{\lambda\sigma}
\eeq
with $ F^a_{\mu\nu} = \nabla_\mu A^a_\nu - \nabla_\nu A^a_\mu
+ g f^{abc} A^b_\mu A^c_\nu $
and $a= 1 \dots N^2 -1 $.
On the lattice one can follow a procedure analogous to Wilson's
construction on a hypercubic lattice, with the main
difference that the lattice is now simplicial.
Given a link $ij$ on the lattice
one assigns group element $ U_{ij} $, with each $U$ an
$N \times N$ unitary matrix with determinant equal to one,
and such that $U_{ji} = U_{ij}^{-1} $.
Then with each triangle (plaquettes) $\Delta$ labeled by the three
vertices $ijk$ one associates a product
of three $U$ matrices 
$ U_{\Delta} \; \equiv \; U_{ijk} \; = U_{ij} \, U_{jk} \, U_{ki} $ .
The discrete action is then given by 
\beq
I_{\rm gauge} \; = \; - { 1 \over g^2 } \, \sum_{\Delta} \; V_{\Delta}
\, {c \over A_\Delta^2 } \, 
{\rm Re} \, \left [ \tr ( 1 \, - \, U_{\Delta} ) \right ]
\label{eq:gauge-ac}
\eeq
with $1$ the unit matrix,  $ V_{\Delta} $ the $4$-volume associated
with the plaquettes $\Delta$, $ A_\Delta $ the area of the triangle
(plaquettes) $\Delta$, and $c$ a numerical constant of order one.
One important property of the gauge lattice action of 
Eq.~(\ref{eq:gauge-ac}) is its local invariance under gauge
rotations $g_i$ defined at the lattice vertices,
and for which $U_{ij}$ on the link $ij$ transforms as
\beq
U_{ij} \; \rightarrow \;  g_i \, U_{ij} \, g_j^{-1}
\eeq


Finally one can consider a spin-$3/2$ field.
Of course supergravity in four dimensions naturally contains a spin-$3/2$
gravitino, the supersymmetric partner of the graviton.
In the case of ${\cal N}=1$ supergravity these are the only two
degrees of freedom present.
Consider here a spin-$3/2$ Majorana fermion in four dimensions, which
correspond to self-conjugate Dirac spinors $\psi_\mu$, where the Lorentz index $\mu =1 \dots 4$.
In flat space the action for such a field is given by the Rarita-Schwinger term
\beq
{\cal L}_{RS} \; = - \; \half \,
\epsilon^{\alpha\beta\gamma\delta} \, \psi^T_\alpha \, C \, 
\gamma_5 \, \gamma_\beta \, \partial_\gamma \, \psi_\delta
\label{eq:rs-field}
\eeq
Locally the action is invariant under the gauge transformation
$ \psi_\mu (x) \; \rightarrow \; \psi_\mu (x) \, + \, 
\partial_\mu \, \epsilon  (x) $,
where $\epsilon  (x)$ is an arbitrary local Majorana spinor.
The construction of a suitable lattice action for the spin-$3/2$
particle proceeds in a way
that is rather similar to what one does in the spin-$1/2$ case.
On a simplicial manifold the Rarita-Schwinger
spinor fields $\psi_\mu (s)$ 
and $\bar{\psi}_\mu (s) $ are most naturally placed
at the center of each $d$-simplex $s$.
Like the spin-$1/2$ case, the
construction of a suitable lattice action requires the
introduction of the Lorentz group and its associated
vierbein fields $e_\mu^a (s) $ within each simplex labeled
by $s$.  
Again as in the spinor case vierbeins $e^\mu_a (s_1)$
and $e^\mu_a (s_2)$ in two neighboring simplices
will be related by a matrix ${\bf R} (s_2, s_1)$ such that
\beq
e_a^\mu (s_2) \; = \; R^\mu_{\;\;\nu} (s_2,s_1) \; e_a^\nu (s_1)
\eeq
and whose spinorial representation $\bf S$
was given previously in Eq.~(\ref{eq:spinrot}).
But the new ingredient in the spin-$3/2$ case is
that, besides requiring a spin rotation matrix 
${\bf S}(s_2,s_1)$, now one also needs
the matrix $ R_\mu^\nu (s,s') $ describing the
corresponding 
parallel transport of the Lorentz vector $\psi_\mu (s)$
from a simplex $s_1$ to the neighboring simplex $s_2$. 
An invariant lattice action for a massless spin-$3/2$ particle
takes therefore the form
\beq
I \; = \; - \, \half \!\! \sum_{\rm faces \; f(s s')} \!\!
V( f(s,s')) \,
\epsilon^{\mu\nu\lambda\sigma} \, 
\bar \psi_\mu (s) \, 
{\bf S} ( {\bf R} (s,s') ) \, \gamma_\nu (s') \,
n_\lambda (s,s') \, R_{\;\; \sigma}^\rho (s,s') \,
\psi_\rho (s' )
\eeq
with 
\beq
\bar \psi_\mu (s) \, {\bf S} ( {\bf R} (s,s') ) \, 
\gamma_\nu (s') \,  \psi_\rho (s' )
\; \equiv \; 
\bar \psi_{\mu \, \alpha} (s) \, 
S^{\alpha}_{\; \; \beta} ( {\bf R} (s,s') ) \, 
\gamma^{\;\; \beta}_{\nu \;\; \gamma} (s') \,  
\psi^\gamma_\rho (s' )
\eeq
and the sum $ \sum_{\rm faces \; f(s s')} $ 
extends over all interfaces $f(s,s')$ connecting one
simplex $s$ to a neighboring simplex $s'$.
When compared to the spin-$1/2$ case, the
most important modification is the 
second rotation matrix $ R_{ \;\; \mu}^\nu (s,s') $, which
describes the parallel transport of the fermionic vector 
$ \psi_\mu $ from the site $s$ to the site $s'$,
which is required in order to obtain locally 
a Lorentz scalar contribution to the action.

\section{Alternate Discrete Formulations}

\label{sec:otherlatt}

The simplicial lattice formulation offers a natural way of representing
gravitational degrees in a discrete framework by employing
inherently geometric concepts such as areas, volumes and angles. 
It is possible though to formulate quantum gravity on a flat
hypercubic lattice, in analogy to Wilson's discrete formulation
for gauge theories, by putting the connection centerstage.
In this new set of theories the natural variables are then lattice
versions of the spin connection and the vierbein.
Also, because the spin connection variables appear 
from the very beginning, it is much easier to incorporate fermions later.
Some lattice models have been based on the pure Einstein theory
while others attempt to incorporate higher derivative terms.

Difficult arise when attempting to put quantum gravity on a flat hypercubic
lattice a la Wilson, since it is not entirely clear what the gravity analogue
of the Yang-Mills connection is.
In continuum formulations invariant under the Poincar\'{e} or de Sitter group
the action is invariant under a local extension of the Lorentz transformations,
but not under local translations.
Local translations are replaced by diffeomorphisms which
have a different nature. 
One set of lattice discretizations starts from the action whose local invariance group is the de Sitter group $Spin(4)$, the
covering group of $SO(4)$.
In one lattice formulation the lattice variables are gauge potentials 
$e_{a\mu} (n)$ and $\omega_{\mu ab}(n)$
defined on lattice sites $n$, generating local $Spin(4)$ matrix transformations
with the aid of the de Sitter generators $P_a$ and $M_{ab}$.
The resulting lattice action reduces classically
to the Einstein action with cosmological term in first order form in the limit of the lattice spacing $a \rightarrow 0$; 
to demonstrate the quantum equivalence one needs an additional zero torsion
constraint.
In the end the issue of lattice diffeomorphism invariance remains somewhat open,
with the hope that such an invariance will be restored in the full quantum theory.

As an example, we will discuss here the approach of Mannion and Taylor,
which relies on a four-dimensional lattice discretization of the 
Einstein-Cartan
theory with gauge group $SL(2,C)$, and does not initially require the presence
of a cosmological constant, as would be the case if one had started out
with the de Sitter group $Spin(4)$.
On a lattice of spacing $a$ with vertices labelled by $n$ and directions
by $\mu$ one relates
the relative orientations of nearest-neighbor local $SL(2,C)$ frames by
\beq
U_\mu (n) = \Bigr [ U_{-\mu} ( n+ \mu ) \Bigl ]^{-1} 
\, = \, \exp [ \, i \, B_\mu (n) ]
\eeq
with $B_\mu = \half a B_\mu^{ab} (n) J_{ba}$, $J_{ba}$ being the set of six
generators of $SL(2,C)$, the covering group of the Lorentz group $SO(3,1)$,
usually taken to be 
$ \sigma_{ab} = { 1 \over 2 i} [ \gamma_a , \gamma_b ] $
with $\gamma_a$'s the Dirac gamma matrices.
The local lattice curvature is then obtained in the usual way
by computing the product of four parallel transport matices
around an elementary lattice square,
\beq
U_\mu (n) \, U_\nu (n + \mu) \, U_{-\mu} (n+\mu+\nu) \, U_{-\nu} (n+\nu)
\eeq
giving in the limit of small $a$ by the Baker-Hausdorff formula
the value $\exp [ i a R_{\mu\nu} (n) ] $, where $R_{\mu\nu}$
is the Riemann tensor defined in terms of the spin connection $B_\mu$
\beq
R_{\mu\nu} = \partial_\mu B_\nu - \partial_\nu B_\mu + i [ B_\mu, B_\nu ]
\eeq
If one were to write for the action the usual Wilson lattice gauge form
\beq
\sum_{n, \mu, \nu} \tr [ \,
U_\mu (n) \, U_\nu (n + \mu) \, U_{-\mu} (n+\mu+\nu) \, U_{-\nu} (n+\nu) \, ]
\eeq
then one would obtain a curvature squared action proportional to
$\sim \int R_{\mu\nu}^{\;\;ab} R^{\mu\nu}_{\;\;ab} $
instead of the Einstein-Hilbert one.
One needs therefore to introduce lattice vierbeins $e_\mu^{\;\;b} (n)$
on the sites by defining the matrices
$ E_\mu (n) = a \, e_\mu^{\;\;a} \, \gamma_a $.
Then a suitable lattice action is given by
\beq
I = { i \over 16 \kappa^2 }
\sum_{n, \mu, \nu, \lambda, \sigma} \tr [ \, \gamma_5 \,
U_\mu (n) \, U_\nu (n + \mu) \, U_{-\mu} (n+\mu+\nu) \, U_{-\nu} (n+\nu) \,
E_\sigma (n) \, E_\lambda (n) \, ]
\label{eq:lagr-mt}
\eeq
The latter is invariant under local $SL(2,C)$ transformations $\Lambda (n)$
defined on the lattice vertices
\beq
U_\mu \rightarrow \Lambda (n) \, U_\mu (n) \, \Lambda^{-1} (n+\mu)
\eeq
for which the curvature transforms as
\bea
&& U_\mu (n) \, U_\nu (n + \mu) \, U_{-\mu} (n+\mu+\nu) \, U_{-\nu} (n+\nu)
\nonumber \\
&& \;\;\;\;\;\;\;\; \rightarrow \;
\Lambda (n) \,
U_\mu (n) \, U_\nu (n + \mu) \, U_{-\mu} (n+\mu+\nu) \, U_{-\nu} (n+\nu) \,
\Lambda^{-1} (n)
\eea
and the vierbein matrices as
\beq
E_\mu (n) \rightarrow \Lambda (n) \, E_\mu (n) \, \Lambda^{-1} (n)
\eeq
Since $\Lambda (n) $ commutes with $\gamma_5$, the expression in
Eq.~(\ref{eq:lagr-mt}) is invariant.
The metric is then obtained as usual by
\beq
g_{\mu\nu} (n) = \quarter \tr [ E_\mu (n) \, E_\nu (n) ] \;\; .
\eeq
From the expression for the lattice curvature
$ R_{\mu\nu}^{\;\;ab} $ given above if follows immediately that the lattice action
in the continuum limit becomes
\beq
I \, = \, { a^4 \over 4 \kappa^2 }
\sum_n \epsilon^{\mu\nu\lambda\sigma} \, \epsilon_{abcd} \,
R_{\mu\nu}^{\;\;\;\;ab} (n) \, e^{\;\;c}_\lambda (n) \, e^{\;\;d}_\sigma (n)
\, + \, O( a^6)
\eeq
which is the Einstein action in Cartan form
\beq
I \, = \, { 1 \over 4 \kappa^2 } \int d^4 x \,
\epsilon^{\mu\nu\lambda\sigma} \, \epsilon_{abcd} \,
R_{\mu\nu}^{\;\;\;\;ab} \, e^{\;\;c}_\lambda \, e^{\;\;d}_\sigma
\eeq
with the parameter $\kappa$ identified with the Planck length.
One can add more terms to the action;
in this theory a cosmological term can be represented  by
\beq
\lambda_0 \sum_n \epsilon^{\mu\nu\lambda\sigma}
\tr [ \, \gamma_5 \,
E_\mu (n) \, E_\nu (n) \, E_\sigma (n) \, E_\lambda (n) \, ]
\label{eq:lagr-mt1}
\eeq
Both Eqs.~(\ref{eq:lagr-mt}) and Eq.~(\ref{eq:lagr-mt1}) are locally
$SL(2,C)$ invariant.
The functional integral is then given by
\beq
Z \, = \, \int \prod_{n, \mu}  d B_\mu (n) \, \prod_{n, \sigma}  d E_\sigma (n)
\, \exp \, \Bigl \{ - I ( B,E) \Bigr \}
\eeq
and from it one can then compute suitable quantum averages.
Here $ d B_\mu (n)$ is the Haar measure for $SL(2,C)$;
it is less clear how to choose the integration measure over the
$E_\sigma$'s, and how it should suitably constrained, which
obscures the issue of diffeomorphism invariance in this theory.

There is another way of discretizing gravity, still using largely
geometric concepts as is done in the Regge theory.
In the dynamical triangulation approach due to David
one fixes the edge lengths to
unity, and varies the incidence matrix. 
As a result the volume of each simplex is fixed at
\beq
V_d \; = \; { 1 \over d! } \sqrt{ d+1 \over 2^d } \;\; ,
\eeq
and all dihedral angles are given by the constant value
\beq
\cos \theta_d \; = \; { 1 \over d } 
\eeq
so that for example in four dimensions one has 
$\theta_d = \arccos (1/4) \approx 75.5^{o}$.
Local curvatures are then determined by how many simplices $n_s (h)$
meet on a given hinge,
\beq
\delta (h)  \; = \; 2 \, \pi - n_s (h ) \, \theta_d
\eeq
The action contribution from a single hinge is
therefore from Eq.~(\ref{eq:regge-d}) 
$\delta (h) A(h) = \quarter \sqrt{3} [ 2 \, \pi - n_s (h ) \, \theta_d ]$
with $n_s$ a positive integer.
In this model the local curvatures are inherently discrete, and there
is no equivalent lattice notion of continuous diffeomorphisms, 
or for that matter of
continuous local deformations corresponding, for example, to shear waves.
Indeed it seems rather problematic in this approach to make contact
with the continuum theory, as the model does not contain a metric,
at least not in an explicit way.
This fact has some consequences for the functional
measure, since there is really no clear criterion which could be 
used to restrict it to the form suggested by
invariance arguments, as detailed earlier in the discussion
of the continuum functional integral for gravity.
The hope is that for lattices made of some large number of simplices
one would recover some sort of discrete version of diffeomorphism invariance.
Recent attempts have focused on simulating the Lorentzian case,
but new difficulties arise in this case as it leads in principle to complex 
weights in the functional integral, which
are next to impossible to handle correctly in numerical simulations 
(since the latter generally rely on positive probabilities).
  
Another lattice approach somewhat related to the Regge theory 
described in this review is based on the so-called
spin foam models, which have their origin in an observation found in
Ponzano and Regge  relating the geometry of simplicial lattices
to the asymptotics of Racah angular momentum addition coefficients.
The original concepts were later developed into a 
spin model for gravity based on quantum spin variables attached to lattice links.
In these models representations of $SU(2)$ label edges.
One natural underlying framework for such theories is the
canonical $3+1$ approach to quantum gravity, wherein quantum spin variables
are naturally related to $SU(2)$ spin connections.
Extensions to four dimensions have been attempted, and we
refer the reader to recent reviews of spin foam models.

\section{Lattice Weak Field Expansion and Transverse-Traceless Modes}

\label{sec:latticewfe}

One of the simplest possible problems that can be treated in quantum Regge
gravity is the analysis of small fluctuations about a fixed flat Euclidean simplicial background.
In this case one finds that the lattice graviton propagator in a 
De Donder-like gauge is precisely analogous to the continuum expression.
To compute an expansion of the lattice Regge action 
\beq
I_R \; \propto \; \sum_{\rm hinges} \, \delta (l) \; A (l) 
\eeq
to quadratic order in the lattice weak fields one needs second variations with respect to the edge lengths.
The second variation about flat space is given by  
\beq
\delta^2 I_R \; \propto \; 
\sum_{\rm hinges} \,
\left ( \sum_{\rm edges} { \partial \delta \over \partial l }
\, \delta l \, \right ) \cdot
\left ( \sum_{\rm edges} { \partial A \over \partial l } \, \delta l \, \right )
\eeq
Next a specific lattice structure needs to be chosen as a background geometry.
A natural choice is to use a flat hypercubic lattice, made
rigid by introducing face diagonals, body diagonals and hyperbody diagonals,
which results into a subdivision of each hypercube into $d!$ (here
4!=24) simplices.
This particular subdivision is shown in Fig.5.

\begin{center}
\epsfxsize=4cm
\epsfbox{4dcubei.eps}
\end{center}

\noindent{\small Figure 5.
A four-dimensional hypercube divided up into four-simplices.}
\medskip

\label{fig:4d-cube}

By a simple translation, the whole lattice can then be constructed from
this one elemental hypercube.
Consequently there will be $2^d - 1 = 15$ lattice fields per point,
corresponding to all the edge lengths emanating in the positive lattice
directions from any one vertex. 
Note that the number of degrees per lattice point is slightly
larger than what one would have in the continuum, where the
metric $g_{\mu\nu}(x)$ has $d(d+1)/2=10$ degrees of freedom per
spacetime point $x$ in four dimensions 
(perturbatively, the physical degrees of freedom in the continuum are much less:
$ {1\over 2} d (d+1)-1-d-(d-1)$ = ${1 \over 2} d (d-3) $, for a traceless symmetric tensor, and after imposing gauge conditions).
Thus in four dimensions each lattice hypercube will contain 4 body principals, 
6 face diagonals, 4 body diagonals and one hyperbody diagonal.
Within a given hypercube it is quite convenient to label the coordinates of the
vertices using a binary notation, so that the four body principals
with coordinates $(1,0,0,0)$ … $(0,0,0,1)$ will be labeled
by integers 1,2,4,8, and similarly for the other vertices  
(thus for example the vertex $(0,1,1,0)$, corresponding to a face diagonal
along the second and third Cartesian direction, will be labeled by the
integer 6).

For a given lattice of fixed connectivity, the edge lengths are then
allowed to fluctuate around an equilibrium value $l_i^0 $
\beq
l_i = l_i^0 \;( 1 + \epsilon_i ) 
\eeq
In the case of the hypercubic lattice subdivided into simplices,
the unperturbed edge lengths $ l_i^0 $ take on the values $1,\sqrt{2},\sqrt{3},2$, depending on edge type.
The second variation of the action then reduces to a quadratic form in 
the 15-component small fluctuation vector $ {\bf \epsilon}_n $
\beq
\delta^2 I_R \; \propto \;  \sum_ { m n } \; {\bf \epsilon }_m^T \;
M_{ m n } \; {\bf \epsilon }_n 
\eeq
Here $M$ is the small fluctuation matrix, whose inverse determines the
free lattice graviton propagator, and the indices $m$ and $n$ 
label the sites on the lattice.
But just as in the continuum, $M$ has zero
eigenvalues and cannot therefore be inverted until one supplies an appropriate
gauge condition. 
Specifically, one finds that the matrix $M$ in four dimensions has
four zero modes corresponding to periodic translations of the lattice,
and a fifth zero mode corresponding to periodic fluctuations in the
hyperbody diagonal.
After block-diagonalization it is found that 4 modes completely decouple
and are constrained to vanish, and thus the remaining degrees of freedom
are 10, as in the continuum, where the metric has 10 independent components.
The wrong sign for the conformal mode, which is present in the
continuum, is also reproduced by the lattice propagator.

Due to the locality of the original lattice action, the matrix $M$ can be considered local as well, since it only couples edge fluctuations on neighboring lattice sites. 
In Fourier space one can write for
each of the fifteen displacements $\epsilon_n^{i+j+k+l}$,
defined at the vertex of the hypercube with labels $(i,j,k,l)$,
\beq
\epsilon_n^{i+j+k+l} \; = \;  
( \omega_1 )^i ( \omega_2 )^j ( \omega_4 )^k ( \omega_8 )^l \; \epsilon_n^0
\label{eq:trans}
\eeq
with $\omega_1 = e^{i k_1} $, $\omega_2 = e^{i k_2} $,
$\omega_4 = e^{i k_3} $ and $\omega_8 = e^{i k_4} $
(it will be convenient in the following to use binary notation for
$\omega$ and $\epsilon$, but the regular notation for $k_i$).
Here and in the following we have set the lattice spacing $a$
equal to one.
The remaining dynamics is encoded in the $10 \times 10$ dimensional
matrix $L_\omega \; = \; A_{10}-{1 \over 18} B B^\dagger $.
By a second rotation, here affected by a matrix $T$, it can finally
be brought into the form
\beq
\tilde L_\omega \; = \; T^{\dagger} \, L_\omega \, T \; = \; 
\left [ 8-( \Sigma + \bar \Sigma ) \right ] 
\left ( \matrix{ \half \beta & 0 \cr 0 & I_6 } \right ) \, - \, C^{\dagger} C
\label{eq:lattwfe}
\eeq
with the matrix $\beta$ given by
\beq
\beta \; = \; \half \left ( \matrix{ 
1  & -1 & -1 & -1 \cr
-1 & 1  & -1 & -1 \cr 
-1 & -1 & 1  & -1 \cr 
-1 & -1 & -1 &  1 \cr 
} \right ) 
\label{eq:beta-mat}
\eeq
The other matrix $C$ appearing in the second term is given by
\beq
C = \left ( \matrix{ 
f_1 & 0 & 0 & 0 & \tilde f_2 & \tilde f_4 & 0 & \tilde f_8 & 0 & 0 \cr
0 & f_2 & 0 & 0 & \tilde f_1 & 0 & \tilde f_4 & 0 & \tilde f_8 & 0 \cr
0 & 0 & f_4 & 0 & 0 & \tilde f_1 & \tilde f_2 & 0 & 0 & \tilde f_8 \cr
0 & 0 & 0 & f_8 & 0 & 0 & 0 & \tilde f_1 & \tilde f_2 & \tilde f_4 \cr
} \right )
\label{eq:c-matrix}
\eeq
with $f_i \equiv \omega_i -1 $ and $\tilde f_i \equiv 1- \bar \omega_i $.
Furthermore $\Sigma = \sum_i \omega_i $, and for small momenta one finds
\beq
8-( \Sigma + \bar \Sigma ) \; = \; 8- \sum_{i=1}^4 ( e^{i k_i} + e^{- i k_i } )
\; \sim \; k^2 + O(k^4)
\eeq
which shows that the surviving terms in the lattice action are indeed
quadratic in $k$. 
At this point one is finally ready for a comparison with the continuum
result, namely with the Lagrangian for pure gravity in the weak field limit,
namely
\bea
{\cal L}_{sym} \; = \; &-&
\half \partial_\lambda \, h_{\lambda \mu} \, \partial_\mu h_{\nu\nu}
+ \half \partial_\lambda \, h_{\lambda \mu} \, \partial_\nu h_{\nu\mu}
\nonumber \\
&-& \quarter \partial_\lambda \, h_{\mu \nu} \, \partial_\lambda h_{\mu\nu}
+ \quarter \partial_\lambda \, h_{\mu \mu} \, \partial_\lambda h_{\nu\nu}
\eea
The latter can be conveniently split into two parts, as follows
\beq
{\cal L}_{sym} \; = \; - \half \partial_\lambda \, h_{\alpha\beta}
V_{\alpha\beta\mu\nu} \, \partial_\lambda h_{\mu\nu} \, + \, \half C^2
\eeq
with
\beq
V_{\alpha\beta\mu\nu} \; = \; \half \, \eta_{\alpha\mu} \eta_{\beta\nu}
-\quarter \, \eta_{\alpha\beta} \eta_{\mu\nu}
\eeq
with metric components $11,22,33,44,12,13,14,23,24,34$ more conveniently 
labeled sequentially by integers $1 \dots 10$.
The gauge fixing term $C_\mu$ is
\beq
C_\mu \; = \; \partial_\nu h_{\mu\nu} - \half \partial_\mu h_{\nu\nu}
\label{eq:gauge-fix-1}
\eeq
The above expression is still not quite the same as the lattice weak
field action, but a simple transformation to trace reversed variables
$\bar h_{\mu\nu} \equiv h_{\mu\nu} - \half \delta_{\mu\nu} h_{\lambda\lambda}$
leads to
\beq
{\cal L}_{sym} \; = \;
\half k_\lambda \bar h_i V_{ij} k_\lambda \bar h_j
\, - \, \half \bar h_i ( C^{\dagger} C )_{ij} \bar h_j
\eeq
with the matrix $V$ given by 
\beq
V_{ij} \; = \; 
\left ( \matrix{ \half \beta & 0 \cr 0 & I_6 } \right )
\eeq
with $k = i \partial $. 
Now $\beta$ is the same as the matrix in Eq.~(\ref{eq:beta-mat}),
and $C$ is nothing but the small $k$ limit of the matrix by the same name
in Eq.~(\ref{eq:c-matrix}). 

It is easy to see that the sequence of transformations expressed by 
the matrices $S$ and $T$ relating the lattice fluctuations 
$\epsilon_i (n) $ to their continuum counterparts $h_{\mu\nu}(x)$,
just reproduces the expected relationship between lattice and continuum
fields.
On the one hand one has $ g_{\mu\nu} = \eta_{\mu\nu} + h_{\mu\nu} $, where
$ \eta_{\mu\nu} $ is the flat metric.
At the same time one has from Eq.~(\ref{eq:latmet}) for each simplex
within a given hypercube
\beq
g_{ij} \; = \; \half ( l_{0i}^2 + l_{0j}^2-l_{ij}^2 )
\eeq
By inserting $l_i \; = \; l_i^0 \; ( 1 + \epsilon_i )$,
with $l_i^0 = 1, \sqrt{2}, \sqrt{3},2 $ for the body principal
($i=1,2,4,8$), face diagonal ($i=3,5,6,9,10,12$), 
body diagonal ($i=7,11,13,14$) and hyperbody diagonal ($i=15$),
respectively, one gets for example $( 1 + \epsilon_1 )^2 = 1 + h_{11}$,
$ ( 1 + \epsilon_3 )^2 = 1 + \half \, ( h_{11} + h_{22} ) + h_{12}$ etc.,
which in turn can then be solved for the $\epsilon$'s in terms
of the $h_{\mu\nu}$'s,
\bea
\epsilon_1 & = & \half \, h_{11} + O( h^2) 
\nonumber \\
\epsilon_3 & = & \half \, h_{12} + \quarter \, ( h_{11} +  h_{22} ) + O( h^2)  
\nonumber \\
\epsilon_7 & = & \sixth \, ( h_{12} + h_{13} + h_{23} ) +
\sixth \, ( h_{23} +  h_{13} + h_{12} )
\nonumber \\
&& + \sixth \, ( h_{11} +  h_{22} + h_{33} ) + O( h^2)  
\nonumber \\
\eea
and so on.
As expected, the lattice action has a local gauge invariance, whose explicit
form in the weak field limit can be obtained explicitly.
This continuous local invariance has $d$ parameters in $d$ dimensions and
describes therefore lattice diffeomorphisms.
In the quantum theory, such local gauge invariance
implies the existence of Ward identities for $n$-point functions.

\section{Strong Coupling Expansion}

\label{sec:strong}

In this section the strong coupling
(large $G$ or small $k=1/(8 \pi G)$)
expansion of the lattice gravitational functional integral will be discussed.
The resulting series is in general expected to be useful up to some $k=k_c$,
where $k_c$ is the lattice critical point,
at which the partition function develops a singularity.
One starts from the lattice regularized
path integral with action Eq.~(\ref{eq:latac}) and 
measure Eq.~(\ref{eq:lattmeas}).
Then the four-dimensional Euclidean lattice action contains
the usual cosmological constant and Regge scalar curvature terms
\beq 
I_{latt} \; = \;  \lambda \, \sum_h V_h (l^2) \, - \, 
k \sum_h \delta_h (l^2 ) \, A_h (l^2) \;\; , 
\label{eq:ilatt1} 
\eeq
with $k=1/(8 \pi G)$, and possibly additional higher derivative terms
as well.
The action only couples edges which belong either to
the same simplex or to a set of neighboring simplices, and can therefore
be considered as {\it local}, just like the continuum action. 
It leads to a lattice partition function defined in Eq.~(\ref{eq:zlatt})
\beq 
Z_{latt} \; = \;  \int [ d \, l^2 ] \; e^{ 
- \lambda_0 \sum_h V_h \, + \, k \sum_h \delta_h A_h } \;\; ,
\label{eq:zlatt-1} 
\eeq
where, as customary, the lattice ultraviolet cutoff is set equal to one
(i.e. all length scales are measured in units of the lattice cutoff).
For definiteness the measure will be of the form 
\beq
\int [ d \, l^2 ] \; = \;
\int_0^\infty \; \prod_s \; \left ( V_d (s) \right )^{\sigma} \;
\prod_{ ij } \, dl_{ij}^2 \; \Theta [l_{ij}^2] \;\; .
\eeq
When doing an expansion in the kinetic term
proportional to $k$, it is convenient to include the
$\lambda$-term in the measure. 
We will set therefore here as in Eq.~(\ref{eq:mulatt})
\beq 
d \mu (l^2) \; \equiv \; [ d \, l^2 ] \, e^{- \lambda_0 \sum_h V_h }
\;\; .
\label{eq:mulatt1} 
\eeq
As a next step, $Z_{latt}$ is expanded in powers of $k$,
\beq 
Z_{latt}(k) \; = \;  \int d \mu (l^2) \, \; e^{k \sum_h \delta_h \, A_h } 
\; = \;  \sum_{n=0}^{\infty} \, { 1 \over n!} \, k^n \, 
\int d \mu (l^2) \, \left ( \sum_h \delta_h \, A_h \right )^n \;\; .
\label{eq:zlatt-k}
\eeq
It is easy to show that $Z (k) \, = \, \sum_{n=0}^{\infty} a_n \, k^n $
is analytic at $k=0$, so this expansion should be well defined up to the
nearest singularity in the complex $k$ plane.
One key quantity in the strong coupling expansion of lattice gravity is the
{\it correlation} between different plaquettes,
\beq
< ( \delta \, A )_{h} \, ( \delta \, A)_{h'} > \; = \; 
{\displaystyle 
\int d \mu (l^2) \,
( \delta \, A)_{h} \, ( \delta \, A)_{h'} \,
e^{k \sum_h \delta_h \, A_h }
\over
\displaystyle \int d \mu (l^2) \, e^{k \sum_h \delta_h \, A_h } } \;\; ,
\label{eq:corr}
\eeq
or, better, its {\it connected} part (denoted here by $< \dots >_C$).
Here again the exponentials in the numerator and denominator can be
expanded out in powers of $k$.
The lowest order term in $k$ will involve the correlation
\beq
\int d \mu (l^2) \, ( \delta \, A)_{h} \, ( \delta \, A)_{h'} \;\; .
\eeq
But unless the two hinges are close to each other, they will fluctuate
in an uncorrelated manner, with
$< ( \delta \, A )_{h} \, ( \delta \, A)_{h'} > -
< ( \delta \, A )_{h} > < ( \delta \, A)_{h'} > \, = \, 0 $.
In order to achieve a non-trivial correlation, the path between
the two hinges 
$h$ and $h'$ needs to be tiled by at least as many terms from
the product $ ( \sum_h \delta_h \, A_h )^n $ in
\beq
\int d \mu (l^2) \, ( \delta \, A)_{h} \, ( \delta \, A)_{h'} \,
\left ( \sum_h \delta_h \, A_h \right )^n
\eeq
as are needed to cover the distance $l$ between the two hinges.
One then has
\beq
< ( \delta \, A )_{h} \, ( \delta \, A)_{h'} >_C \; \sim \; 
k^l \; \sim \; e^{- l / \xi } \;\; ,
\label{eq:kxi}
\eeq
with the correlation length $ \xi = 1 / | \log k | \rightarrow 0 $
to lowest order as $k \rightarrow 0 $
(here we have used the usual definition of the correlation length $\xi$,
namely that a generic correlation function
is expected to decay as $ \exp (- {\rm distance} / \xi) $ for
large separations).
This last result is quite general, and holds for example irrespective
of the boundary conditions (unless of course $\xi \sim L$, where $L$ is the
linear size of the system, in which case a path can be found
which wraps around the lattice).
But further thought reveals that the above result is in fact not
completely correct, due to the fact that in order to achieve
a non-vanishing correlation one needs, at least to lowest order,
to connect the two hinges by a narrow tube.
The previous result should then read correctly as 
\beq
< ( \delta \, A )_{h} \, ( \delta \, A)_{h'} >_C \; \sim \; 
\left ( k^{n_d} \right )^l \;\; ,
\label{eq:nd}
\eeq
where $n_d \, l$ represents the minimal number of dual
lattice polygons needed to form a closed surface connecting
the hinges $h$ and $h'$, with $l$ the actual distance (in lattice units) 
between the two hinges.
Figure 6. provides an illustration of the situation.

\begin{center}
\epsfxsize=8cm
\epsfbox{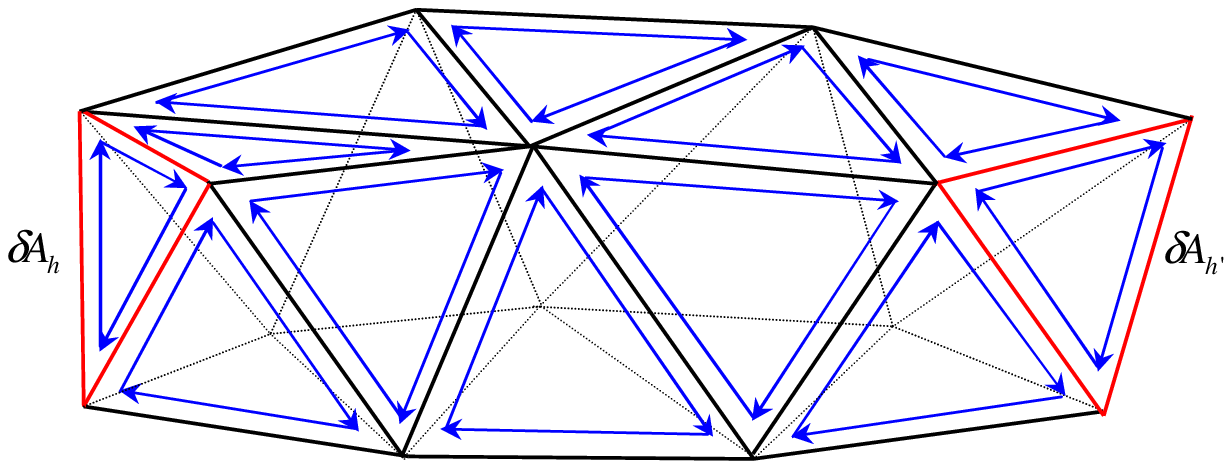}
\end{center}

\noindent{\small Figure 6.
Correlations between action contributions on hinge $h$
and hinge $h'$ arise to lowest order in the strong coupling expansions
from diagrams describing a narrow tube connecting the two hinges.
Here vertices represent points in the dual lattice, with
the tube-like closed surface tiled with parallel transport polygons.
For each link of the dual lattice, the $SO(4)$ parallel transport matrices
${\bf R}$ are represented by an arrow.}
\medskip

\label{fig:tube}

With some additional effort many additional terms can be computed in the strong
coupling expansion.
In practice the method is generally not really competitive with direct
numerical evaluation of the path  integral via Monte Carlo methods.
But it does provide a new way of looking at the functional
integral, and provide the basis for new approaches, such
as the large $d$ limit to be discussed in the
second half of the next section.

\section{Gravitational Wilson Loop}
\label{sec:wilson}

An important question for any theory of quantum gravity is what gravitational
observables should look like, i.e. which expectation values of operators 
(or ratios thereof) have meaning and physical interpretation in the context of 
a manifestly covariant formulation, in particular in a situation where metric 
fluctuations are not necessarily bounded. 
Such averages naturally include the previously discussed
expectation values of the (integrated) scalar curvature and other related 
quantities (involving for example curvature-squared terms), as well as 
correlations of operators at fixed geodesic distance. 
Another set of 
physical observables on which we focus here corresponds to the gravitational analog of the Wilson 
loop.
It provides information about the parallel transport of vectors, and 
therefore on the effective curvature, around large, near-planar loops.
In contrast to gauge theories, the Wilson loop in quantum gravity
does not give information on the static potential, which
is obtained instead for the correlation between particle world-lines.

The gauge theory definition can be adapted to the lattice
gravitational case.
It turns out that it is most easily achieved by using a slight variant of Regge calculus, 
in which the action coincides with the usual Regge action in the near-flat 
limit. 
Here we will use extensively the notion of lattice parallel transport
discussed earlier, and how areas are defined on the dual lattice.

At strong coupling the measure and cosmological constant terms
form the dominant part of the functional integral, since the Einstein
part of the action is vanishingly small in this limit.
Yet, and in contrast to strongly coupled lattice Yang-Mills theories,
the functional integral is still non-trivial to compute analytically in
this limit, mainly due to the triangle inequality constraints.
Therefore, in order to be able to derive some analytical estimates for
correlation functions in the strong coupling limit, one needs still
to develop some set of approximation methods.
In principle the reliability of the approximations can later be
tested by numerical means,
for example by integrating directly over edges using
the explicit  lattice measure given above.

One approach that appears natural in the gravity context
follows along the lines of what is normally done in gauge
theories, namely an integration over compact group variables,
using the invariant measure over the gauge group.
It is of this method that we wish to take advantage here, as we
believe that it is well suited for gravity as well.
In order to apply such a technique to gravity one needs (i) to
formulate the lattice theory in such a way that group variables are separated and
therefore appear
explicitly; (ii) integrate over the group variables using an invariant
measure; and (iii) approximate the relevant 
correlation functions in such a way
that the group integration can be performed exactly, using for example
mean field methods for the parts that appear less tractable.
In such a program one is aided by the fact that
in the strong coupling limit one is expanding about a well defined
ground state, and that the measure and the interactions are
{\it local}, coupling only lattice variable (edges or rotations) 
which are a few lattice spacings apart.
The downside of such methods is that one is no longer evaluating the 
functional integral for quantum gravity exactly, even in the strong
coupling limit; the upside is that one obtains a clear analytical
estimate, which later can be in principle systematically tested by 
numerical methods (which are in principle exact).

In the gravity case the analogs of the gauge variables of Yang-Mills
theories are given by the connection, so it is natural therefore
to look for a first order formulation of Regge gravity.
The main feature of this approach is that one treats the 
metric $g_{\mu\nu}$ and the affine connection 
$\Gamma_{ \mu \nu }^\lambda$ as independent variables.
There one can safely consider functionally integrating
separately over the affine connection and the metric, treated
as independent variables, with the correct relationship between
metric and connection arising then as a consequence of the dynamics.
In the lattice theory we will follow a similar spirit, separating
out explicitly in the lattice action the degrees of freedom
corresponding to local rotations 
(the analogs of the $\Gamma$'s in the
continuum), which we will find to be most conveniently described
by orthogonal matrices ${\bf R}$.

The next step is a use of the properties of local rotation 
matrices in the
context of the lattice theory, and how these
relate to the lattice gravitational action.
It was shown earlier that with each neighboring pair of simplices $s,s+1$ one 
can associate a
Lorentz transformation $ R^{\mu}_{\;\; \nu} (s,s+1)$.
For a closed elementary path 
$C_h$ encircling a hinge $h$ and passing through
each of the simplices that meet at that hinge one has
for the total rotation matrix ${\bf R} \equiv \prod_s  R_{s,s+1} $
associated with the given hinge 
\beq
\Bigl [ \prod_s  R_{s,s+1}   \Bigr ]^{\mu}_{\;\; \nu} \; = \;
\Bigl [ \, e^{\delta (h) U (h)} \Bigr ]^{\mu}_{\;\; \nu}  \;\; ,
\label{eq:wi-fullrot}
\eeq
\noindent as in Eq.~(\ref{eq:rot-hinge}). 
More generally one might want to consider a near-planar, but non-infinitesimal,
closed loop $C$, as shown in Fig.7.
Along this closed loop the overall rotation matrix will still be given by 
\beq
R^{\mu}_{\;\; \nu} (C) \; = \;
\Bigl [ \prod_{s \, \subset C}  R_{s,s+1} \Bigr ]^{\mu}_{\;\; \nu} 
\eeq
In analogy with the infinitesimal loop case,
one would like to state that for the overall rotation matrix one has
\beq
R^{\mu}_{\;\; \nu} (C) \; \approx \; 
\Bigl [ \, e^{\delta (C) U (C))} \Bigr ]^{\mu}_{\;\; \nu}  \;\; ,
\label{eq:latt-wloop-a}
\eeq
where $U_{\mu\nu} (C)$ is now an area bivector perpendicular to the
loop, which will work only if the loop is close to planar so
that $U_{\mu\nu}$ can be taken to be approximately constant
along the path $C$. By a near-planar loop around the point $P$, we mean 
one that is constructed by drawing outgoing geodesics, on a plane through $P$.

\begin{center}
\epsfxsize=8cm
\epsfbox{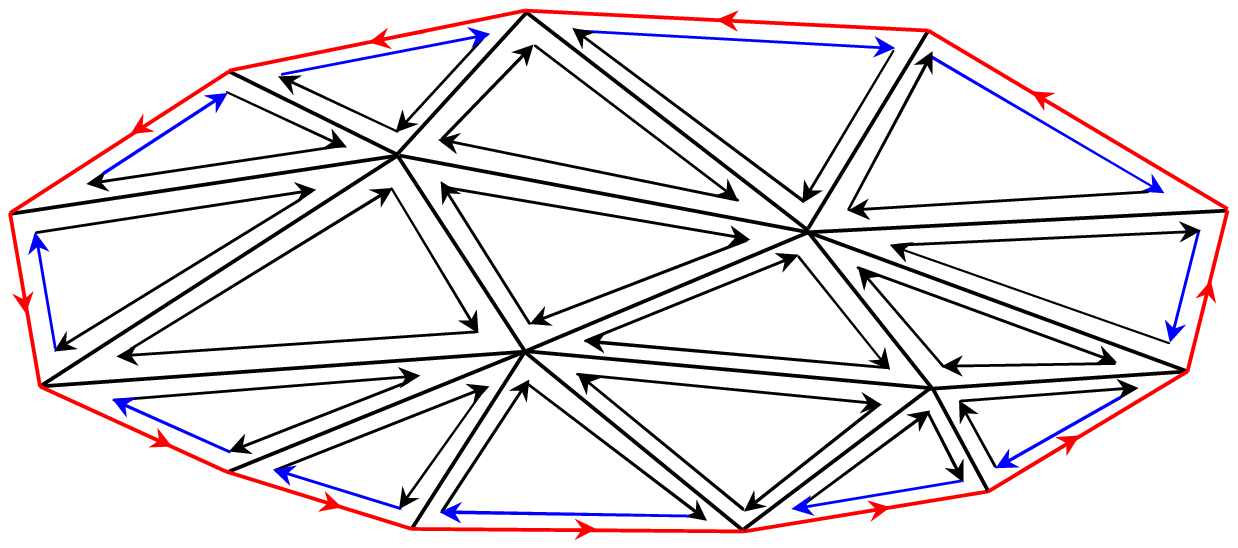}
\end{center}

\noindent{\small Figure 7.
Gravitational analog of the Wilson loop.
A vector is parallel-transported along the larger outer loop.
The enclosed minimal surface is tiled with parallel
transport polygons, here chosen to be triangles for illustrative
purposes.
For each link of the dual lattice, the elementary parallel transport
matrices ${\bf R}(s,s')$ are represented by arrows. 
In spite of the fact that the (Lorentz) matrices ${\bf R}$ can fluctuate 
strongly in accordance with the local geometry, two contiguous, oppositely
oriented arrows always give ${\bf R} {\bf R}^{-1} = 1$.}
\medskip

\label{fig:wilson}

If that is true, then one can define an appropriate coordinate scalar 
by contracting the above rotation matrix ${\bf R}(C)$ 
with the some appropriate bivector, namely
\beq
W ( C ) \; = \; \omega_{\alpha\beta}(C) \, R^{\alpha\beta} (C) 
\label{eq:latt-wloop1-a}
\eeq
where the bivector,
$\omega_{\alpha\beta} (C )$, is intended as being representative of the overall
geometric features of the loop
(for example, it can be taken as an average of the hinge bivector
$\omega_{\alpha\beta} (h)$ along the loop).

In the quantum theory one is interested in the average
of the above loop operator $W (C)$.
Now one notes that for any continuum manifold one can define locally
the parallel transport of a vector around a near-planar loop
$C$.
If the curvature of the manifold is small, one can treat the larger loop
the same way as the small one; then the expression of 
Eq.~(\ref{eq:latt-wloop-a})
for the rotation matrix ${\bf R} (C )$ associated with a near-planar
loop can be re-written in terms of a surface
integral of the large-scale Riemann tensor, projected along the surface
area element bivector $A^{\alpha\beta} (C )$ associated with the loop,
\beq
R^{\mu}_{\;\; \nu} (C) \; \approx \; 
\Bigl [ \, e^{\half \int_S 
R^{\, \cdot}_{\;\; \cdot \, \alpha\beta} \, A^{\alpha\beta} ( C )} 
\Bigr ]^{\mu}_{\;\; \nu}  \;\; .
\eeq
Thus a direct calculation of the Wilson loop provides a way of determining
the {\it effective} curvature at large distance scales, even in the case
where short distance fluctuations in the metric may be significant.

A detailed lattice calculation at strong coupling then gives the following result.
First one dfines the lattice Wilson loop as
\beq
 W(C)  \; = \; < \; Tr[(U_C \; + \; \epsilon \; I_4) \; 
R_1 \; R_2 \; ... \; ... \; R_n] \; > \; .
\label{eq:wloop-def2}
\eeq
where the $R_i$'s are the rotation matrices along the path and the factor 
$(U_C + \epsilon I_4)$, containing some \lq\lq average" direction bivector, 
$U_C$, for the loop, which, after all, is assumed to be almost planar.
For sufficiently strong coupling one obtains an area law, in other words
the above quantity behaves for large areas as
\beq
\exp [ \,( A_C / {\bar A}) \, \log (k \, {\bar A} / {16}) \, ] 
\, = \, \exp ( - \, A_C / {\xi }^2 )
\label{eq:expdecay}
\eeq
where 
$\xi \equiv [{\bar A} / \vert \log (k \, {\bar A} / {16}) \vert]^{1/2} $. 
The rapid decay of the quantum gravitational Wilson loop as a function of the
area is seen as a general and direct consequence of the
assumed disorder in the uncorrelated fluctuations of the parallel transport
matrices ${\bf R}(s,s')$ at strong coupling.

Here it is important to note that the gravitational correlation length $\xi$ is defined 
independently of the expectation value of the Wilson loop.
Indeed a key quantity in gauge theories as well as gravity is the
{\it correlation} between different plaquettes, which in simplicial 
gravity is given by (see Eq.~(\ref{eq:corr})),
\beq
< ( \delta \, A )_{h} \, ( \delta \, A)_{h'} > \; = \; {\displaystyle
\int d \mu (l^2) \,
( \delta \, A)_{h} \, ( \delta \, A)_{h'} \,
e^{k \sum_h \delta_h \, A_h }
\over
\displaystyle \int d \mu (l^2) \, e^{k \sum_h \delta_h \, A_h } } \;\; .
\label{eq:corr-1}
\eeq

The final step is an interpretation of this last 
result in semi-classical terms.
As discussed at the beginning of this section,
the rotation matrix appearing in the gravitational Wilson loop can be related classically to a well-defined physical process:
a vector is parallel transported around a large loop, and at the end it
is compared to its original orientation.
The vector's rotation is then directly related to some sort
of average curvature enclosed by the loop.
The total rotation matrix ${\bf R}(C)$ is given in general by a
path-ordered (${\cal P}$) exponential of the integral of the
affine connection $ \Gamma^{\lambda}_{\mu \nu}$ via
\beq
R^\alpha_{\;\; \beta} (C) \; = \; \Bigl [ \; {\cal P} \, \exp
\left \{ \oint_
{{\bf path \; C}}
\Gamma^{\cdot}_{\lambda \, \cdot} d x^\lambda
\right \}
\, \Bigr ]^\alpha_{\;\; \beta}  \;\; .
\label{eq:rot-cont-a}
\eeq
In such a semi classical description of the parallel transport
process of a vector around a very large loop, one can re-express
the connection in terms of a suitable coarse-grained, or semi-classical, 
Riemann tensor, using Stokes' theorem
\beq
R^\alpha_{\;\; \beta} (C) \; \sim \;
\Bigl [ \;
\exp \, \left \{ \half \,
\int_{S(C)}\, R^{\, \cdot}_{\;\; \cdot \, \mu\nu} \, A^{\mu\nu}_{C} \;
\right \}
\, \Bigr ]^\alpha_{\;\; \beta}  \;\; ,
\label{eq:rot-cont1}
\eeq
where here $ A^{\mu\nu}_{C}$ is the usual area bivector associated with the
loop in question.
The use of semi-classical arguments in relating the above rotation matrix
${\bf R}(C)$
to the surface integral of the Riemann tensor assumes (as usual in the
classical context) that the curvature is slowly varying on the scale
of the very large loop.
Since the rotation is small for weak curvatures, one can write
\beq
R^\alpha_{\;\; \beta} (C) \; \sim \;
\Bigl [ \, 1 \, + \, \half \,
\int_{S(C)}\, R^{\, \cdot}_{\;\; \cdot \, \mu\nu} \, A^{\mu\nu}_{C}
\, + \, \dots \, \Bigr ]^\alpha_{\;\; \beta}  \;\; .
\label{eq:rot-cont2}
\eeq
At this stage one is ready to compare the above expression to the
quantum result
of Eq.~(\ref{eq:expdecay}).
Since one expression [Eq.~(\ref{eq:rot-cont2})] is a 
matrix and the other [Eq.~(\ref{eq:expdecay})] is a scalar, 
we shall take the trace after first 
contracting the rotation matrix with $(U_C \, + \, \epsilon \, I_4)$, as in the 
definition of the Wilson loop, giving
\beq
W(C) \, \sim \, \Tr \left ( (U_C \, + \, \epsilon \, I_4) \, \exp \, 
\left \{ \, \half \,
\int_{S(C)}\, R^{\, \cdot}_{\;\; \cdot \, \mu\nu} \, A^{\mu\nu}_{C} \; 
\right \} \right ) .
\label{eq:wloop_curv1}
\eeq
For the lattice analog of a background manifold with
constant or near-constant large scale curvature one has
\bea
R_{\mu\nu\lambda\sigma} & = & \third \, \lambda \, ( 
g_{\mu\nu} \, g_{\lambda\sigma} \, - \, 
g_{\mu\lambda} \, g_{\nu\sigma} )
\nonumber \\
R_{\mu\nu\lambda\sigma} \, R^{\mu\nu\lambda\sigma} 
& = & {\textstyle { 8 \over 3 } \displaystyle}
\lambda^2
\eea
so that here one can set
\beq 
 R^{\, \alpha}_{\;\; \beta \, \mu\nu} \; = \; 
{\bar R} \; U^{\, \alpha}_{\;\; \beta} \; U_{\mu\nu} ,
\eeq
where ${\bar R}$ is some average curvature over the loop, and the 
$U$'s here will be taken to coincide with $U_C$. 
The trace of the product of 
$(U_C \, + \, \epsilon \, I_4)$ with this expression gives
\beq
Tr( {\bar R} \; U_C^2 \; A_C ) \; = \; - \; 2 \; {\bar R} \; A_C ,
\eeq
where one has used $ U_{\mu\nu} \, A^{\mu\nu}_{C} \, = 2 \, A_C$
(the choice of direction for the bivectors will be such that the latter
is true for all loops).
This is 
to be compared with the linear term from the other exponential expression, 
$- \, A_C / \xi^2 $. Thus the average curvature is computed to be of the 
order 
\beq
{\bar R} \sim 1 / \xi^2 
\eeq 
at least in the small $k =1 / 8 \pi G $ limit.
An equivalent way of phrasing the last result is that $1 / \xi^2$ should
be identified, up to a constant of proportionality, with the
scaled cosmological constant $\lambda$.

\section{Nonperturbative Gravity}

\label{sec:numerical}

The exact evaluation of the lattice functional integral
for quantum gravity by numerical methods
allows one to investigate a regime
which is generally inacessible by perturbation theory, where
the coupling $G$ is strong and quantum fluctutations in the metric
are expected to be large.
The hope in the end is to make contact with the analytic
results obtained, for example, in the $2+\epsilon$
expansion, and determine which scenarios are physically
realized in the lattice regularized model, and then
perhaps even in the real world.

Specifically, one can enumerate several major questions
that one would like to get at least partially answered.
The first one is: which scenarios suggested by perturbation
theory are realized in the lattice theory?
Perhaps a stable ground state for the quantum theory cannot be found, which
would imply that the regulated theory is still inherently
pathological.
Furthermore, if a stable ground state exists for some range
of bare parameters, does it require the
inclusion of higher derivative couplings in an essential
way, or is the minimal theory, with an
Einstein and a cosmological term, sufficient? 
Does the presence of dynamical matter, say in the form
of a massless scalar field, play an important role, or
is the non-perturbative dynamics of gravity
determined largely by the pure gravity sector (as in Yang-Mills
theories)?

More generally, is there any indication that the non-trivial 
ultraviolet fixed
point scenario is realized
in the lattice theory in four dimensions? This would imply, 
as in the non-linear sigma model, the existence of at least
two physically distinct phases and non-trivial exponents.
Which quantity can be used as an order parameter to physically 
describe, in a {\it qualitative}, way the two phases?
A clear physical characterization of the two phases would
allow one, at least in principle, to decide which phase, 
if any, could be realized in nature.
Ultimately this might or might not be possible based on
purely qualitative aspects.
As will discussed below, the lattice continuum limit
is taken in the vicinity of the fixed point,
so close to it is the physically most relevant regime.
At the next level one would hope to be able to establish
a {\it quantitative} connection with those continuum
perturbative results
which are not affected by uncontrollable errors, such
as for example the $2+\epsilon$ expansion discussed earlier.
Since the lattice cutoff and the method of dimensional regularization
cut the theory off in the ultraviolet in rather different
ways, one needs to compare universal quantitities
which are {\it cutoff-independent}.
One example is the critical exponent $\nu$, as well as any other
non-trivial scaling dimension that might arise.
Within the $2+\epsilon$ expansion only {\it one}
such exponent appears, to {\it all} orders in the loop
expansion, as $ \nu^{-1} = - \beta ' (G_c) $.
Therefore one central issue in the lattice regularized theory 
is the value of the universal exponent $\nu$.

Knowledge of $\nu$ would allow one to be more
specific about the running of the gravitational
coupling.
One purpose of the earlier discussion was
to convince the reader that the exponent $\nu$
determines the renormalization group running
of $G (\mu^2)$ in the vicinity of the fixed
point, as in Eq.~(\ref{eq:grun-cont1}) for quantized gravity.
From a practical point of view, on the lattice it is 
difficult to determine the running of 
$G(\mu^2)$ directly from correlation functions ,
since the effects from the running of $G$ are
generally small.
Instead one would like to make use of the analog
of Eqs.~(\ref{eq:m-cont1}) for gravity to determine $\nu$, 
and from there the running of $G$.
But the correlation length $\xi = m^{-1}$ is also difficult
to compute, since it enters the curvature 
correlations at fixed geodesic distance, which are hard
to compute for (genuinely geometric) reasons to be discussed later.
Furthermore, these generally decay exponentially in the distance
at strong $G$, and can therefore be difficult to
compute due to the signal to noise problem of
numerical simulations.
Fortunately the exponent $\nu$ can be determined instead, 
and with good accuracy, from singularities of the derivatives
of the path integral $Z$, whose singular part is
expected, on the basis of very general arguments,
to behave in the vicinity of the fixed point
as $F \equiv - {1 \over V} \ln Z \sim \xi^{-d} $
where $\xi$ is the gravitational correlation length.
From Eq.~(\ref{eq:m-cont1}) relating $\xi (G) $ to
$G-G_c$ and $\nu$ one can then determine $\nu$,
as well as the critical coupling $G_c$.

The starting point is once again the lattice regularized
path integral with action as in Eq.~(\ref{eq:latac}) and
measure as in Eq.~(\ref{eq:lattmeas}),
\beq 
Z_{latt} \; = \;  \int [ d \, l^2 ] \; e^{ 
- \lambda_0 \sum_h V_h \, + \, k \sum_h \delta_h A_h } \;\; ,
\eeq
where, as customary, the lattice ultraviolet cutoff is set equal to one
(i.e. all length scales are measured in units of the lattice cutoff).
The lattice measure is given in Eq.~(\ref{eq:lattmeas}) and
is therefore of the form 
\beq
\int [ d \, l^2 ] \; = \;
\int_0^\infty \; \prod_s \; \left ( V_d (s) \right )^{\sigma} \;
\prod_{ ij } \, dl_{ij}^2 \; \Theta [l_{ij}^2] \;\; .
\eeq
with $\sigma$ a real parameter.
Ultimately the above lattice partition function $Z_{latt}$ is intended as a
regularized form of the continuum Euclidean Feynman path integral
of Eq.~(\ref{eq:zcont}).


Among the simplest quantum mechanical averages is the
one associated with the local curvature
\beq
{\cal R} (k) \; \sim \;
{ < \int d x \, \sqrt{ g } \, R(x) >
\over < \int d x \, \sqrt{ g } > } \;\;\; ,
\eeq
The curvature associated with the quantity above is the
one that would be detected when parallel-tranporting 
vectors around infinitesimal loops, with size
comparable to the average lattice spacing $l_0$.
Closely related to it is the fluctuation
in the local curvature 
\beq
\chi_{\cal R}  (k) \; \sim \;
{ < ( \int dx \sqrt{g} \, R )^2 > - < \int dx \sqrt{g} \, R >^2 
\over < \int dx \sqrt{g} > } \;\;\; .
\eeq
The latter is related to the connected curvature correlation at
zero momentum
\beq
\chi_{\cal R} \; 
\sim \; { \int dx \int dy < \sqrt{g(x)} R(x) \; \sqrt{g(y)} R(y) >_c
\over < \int dx \sqrt{g(x)} > } \;\;\; .
\eeq
Both ${\cal R}(k)$ and $\chi_{\cal R}(k)$ are directly
related to derivatives of $Z$ with respect to $k$,
\beq
{\cal R} (k) \, \sim \,
\frac{1}{V} \frac{\partial}{\partial k} \ln Z  \; ,
\label{eq:avrz} 
\eeq
and 
\beq
\chi_{\cal R}  (k) \, \sim \,
\frac{1}{V} \frac{\partial^2}{\partial k^2} \ln Z \; .
\label{eq:chirz} 
\eeq
Thus a divergence or non-analyticity in $Z$, as caused for example
by a phase transition, is expected to show up in these local averages as well.
Note that the above expectation values are manifestly invariant, since
they are related to derivatives of $Z$.


When computing correlations, new issues arise in quantum gravity
due to the fact that the physical distance between any two points $x$ and $y$
\beq
d(x,y \, \vert \, g) \; = \; \min_{\xi} \; \int_{\tau(x)}^{\tau(y)} d \tau \;
\sqrt{ \textstyle g_{\mu\nu} ( \xi )
{d \xi^{\mu} \over d \tau} {d \xi^{\nu} \over d \tau} \displaystyle } \;\; ,
\eeq
is a fluctuating function of the background metric $g_{\mu\nu}(x)$.
In addition, the Lorentz group used to classify spin states is
meaningful only as a local concept. 
In the continuum the shortest distance between two events is determined
by solving the geodesic equation
\beq
{ d^2 x^{\mu} \over d \tau^2 } \, + \, 
\Gamma^{\mu}_{\lambda\sigma} \, 
{ d x^{\lambda} \over d \tau }
{ d x^{\sigma} \over d \tau } \, = \, 0
\eeq
On the lattice the geodesic distance between two 
lattice vertices $x$ and $y$ requires the determination of
the shortest lattice path connecting several lattice vertices,
and having the two given vertices as endpoints.
This can be done at least in principle
by enumerating all paths connecting the two
points, and then selecting the shortest one.
Consequently physical correlations have to be defined at fixed
geodesic distance $d$, as in the following correlation
between scalar curvatures
\beq
< \int d x \, \int d y \, \sqrt{g} \, R(x) \; \sqrt{g} \, R(y) \;
\delta ( | x - y | - d ) > 
\eeq
Generally these do not go to zero at large separation, so
one needs to define the connected part, by
subtracting out the value at $d=\infty$.
These will be indicated in the following by the connected $<>_c$ average,
and we will write the resulting connected curvature correlation
function at fixed geodesic distance compactly as
\beq
G_R (d) \; \sim \; < \sqrt{g} \; R(x) \; \sqrt{g} \; R(y) \;
\delta ( | x - y | -d ) >_c \; .
\eeq
One can define several more invariant correlation functions at 
fixed geodesic distance for other operators involving curvatures.
Thus one is naturally lead to the
connected correlation function
\beq
G_R (d) \; \equiv \; < \sum_{ h \supset x } \delta_h A_h \;
\sum_{ h' \supset y } \delta_{h'} A_{h'} \;
\delta ( | x - y | -d ) >_c \; ,
\eeq
which probes correlations in the scalar curvatures.

In general one expects for the curvature
correlation either a power law decay, for distances sufficiently
larger than the lattice spacing $l_0$,
\beq
< \sqrt{g} \; R(x) \; \sqrt{g} \; R(y) \; \delta ( | x - y | -d ) >_c
\;\; \mathrel{\mathop\sim_{d \; \gg \; l_0 }} \;\; 
{ 1 \over d^{2 n} }  \;\;\;\; ,
\label{eq:rr-pow1}
\eeq
with $n$ some exponent characterizing the power law decay,
or at very large distances an exponential decay,
characterized by a correlation length $\xi$,
\beq
< \sqrt{g} \; R(x) \; \sqrt{g} \; R(y) \; \delta ( | x - y | -d ) >_c
\; \; \mathrel{\mathop\sim_{d \; \gg \; \xi }} \;\;
e^{-d / \xi } \;\;\;\; .
\label{eq:rr-exp}
\eeq
In practice the correlation functions at fixed geodesic distance
are difficult to compute numerically, and therefore not the best route
to study the critical properties.
But scaling arguments allow one to determine the scaling
behavior of correlation functions from critical exponents
characterizing the singular behavior of the {\it free energy}
and various local averages in the vicinity of the critical point.
In general a divergence of the correlation length $\xi$
\beq
\xi (k) \; \equiv \; 
\mathrel{\mathop\sim_{ k \rightarrow k_c}} \; A_\xi \;
| k_c - k | ^{ -\nu }
\label{eq:xi-k}
\eeq
signals the presence of a phase transition, and leads to the appearance
of a singularity in the free energy $F(k)$.
The scaling assumption for the free energy postulates that a divergent
correlation length in the vicinity of the critical point at $k_c$
leads to non-analyticities of the type
\beq
F \equiv - { 1 \over V } \, \ln Z \; = \; F_{reg} + F_{sing}
\;\;\;\;\;\; F_{sing} \sim \xi^{-d}
\label{eq:fsing}
\eeq
where the second relationship follows simply from dimensional arguments
(the free energy is an extensive quantity).
The regular part $F_{reg}$ is generally not determined from $\xi$
by purely dimensional considerations, but
as the name implies is a regular function in the vicinity
of the critical point.
Combining the definition of $\nu$ in Eq.~(\ref{eq:xi-k}) with
the scaling assumption of Eq.~(\ref{eq:fsing}) one obtains
\beq
F_{sing} (k) \; 
\mathrel{\mathop\sim_{ k \rightarrow k_c}} \; | k_c - k | ^{ d \nu }
\eeq
The presence of a phase transition can then
be inferred from non-analytic terms
in invariant averages, such as the average curvature
and its fluctuation.
Thus for the average curvature one obtains
\beq
{\cal R} (k) \; \mathrel{\mathop\sim_{ k \rightarrow k_c}} \;
A_{\cal R} \, | k_c - k |^{d \nu -1} \;\; ,
\label{eq:rsing}
\eeq
up to regular contributions (i.e. constant terms in the
vicinity of $k_c$).
Similarly one has for the curvature fluctuation
\beq
\chi_{\cal R} (k) \; \mathrel{\mathop\sim_{ k \rightarrow k_c}} \;
A_{ \chi_{\cal R} } \; | k_c - k | ^{ -(2- d \nu) } \;\;\;\; .
\label{eq:chising}
\eeq
At a critical point the fluctuation $\chi$ is in general
expected to diverge, corresponding to the
presence of a divergent correlation length.
From such averages one can therefore in principle extract
the correlation length exponent $\nu$ of 
Eq.~(\ref{eq:xi-k}) without having to compute a correlation
function.


As far as the lattice is concerned, one
starts for example with the 4-d hypercube of Fig.5
divided into simplices, and then stacks a number of such
cubes in such a way as to construct an arbitrarily large lattice,
as shown in Fig.~8.
Other lattice structures are of course possible, including even a
random lattice.
The expectation is that for long range correlations involving
distance scales much larger than the lattice spacing the 
precise structure of the underlying lattice structure
will not matter.
This expectation of the existence of a unique scaling limit
is known as {\it universality of critical behavior}.

\begin{center}
\epsfxsize=8cm
\epsfbox{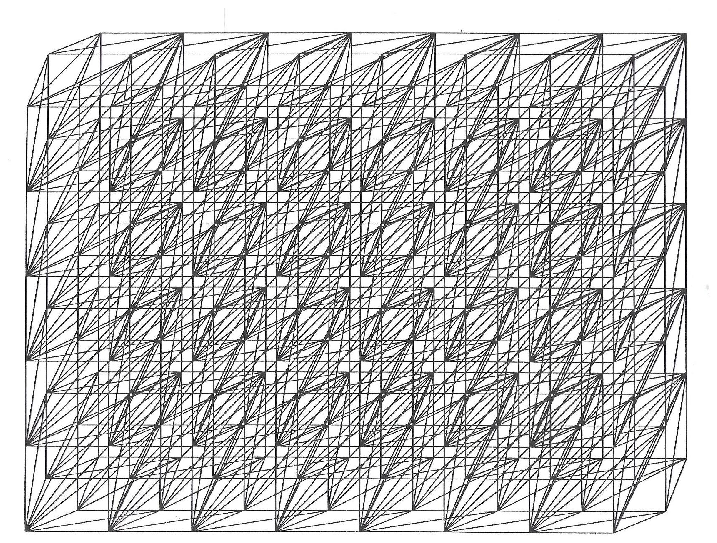}
\end{center}

\noindent{\small Figure 8.
Four-dimensional hypercubes divided into simplices and
stacked to form a four-dimensional lattice.}
\medskip

\label{fig:lat4d}

Typically the lattice sizes investigated range from $4^4$ sites 3840 edges)
to $32^4$ sites (15,728,640 edges).
On a dedicated massively parallel supercomputer millions of 
consecutive edge length configurations can be generated for tens
of values of $k$ in a few day's or week's time.
Furthermore the bare cosmological constant $\lambda_0$ appearing in the
gravitational action of Eq.~(\ref{eq:zlatt}) can be fixed at $1$
in units of the cutoff, since it just sets the overall length
scale in the problem.
The higher derivative coupling $a$ can be set to a value very
close to $0$ since one ultimately is interested in the limit
$a \rightarrow 0$, corresponding to the pure Einstein theory.

One finds that for the measure in Eq.~(\ref{eq:dewlattmeas}) this choice of
parameters leads to a well behaved ground state for $k < k_c $
for higher derivative coupling $a \rightarrow 0$.
The system then resides in the `smooth' phase, with an effective 
dimensionality close to four.
On the other hand for $k > k_c$ the curvature becomes very large
and the lattice collapses into degenerate configurations
with very long, elongated simplices
(see Fig.~9.).

One finds that as $k$ is varied, the average curvature $\cal R$ is
negative for 
sufficiently small $k$ ('smooth' phase), and appears to go to zero 
continuously at some finite value $k_c$.
For $k > k_c$ the curvature becomes very large, and
the simplices tend to collapse into degenerate configurations
with very small volumes ($ <\!V\!> / <\!l^2\!>^2 \sim 0$).
This 'rough' or 'collapsed' phase 
is the region of the usual weak field expansion ($G \rightarrow 0$).
In this phase the lattice collapses into degenerate configurations
with very long, elongated simplices.
This phenomenon is usually intepreted as a lattice remnant of the conformal
mode instability of Euclidean gravity discussed earlier.


\begin{center}
\epsfxsize=10cm
\epsfbox{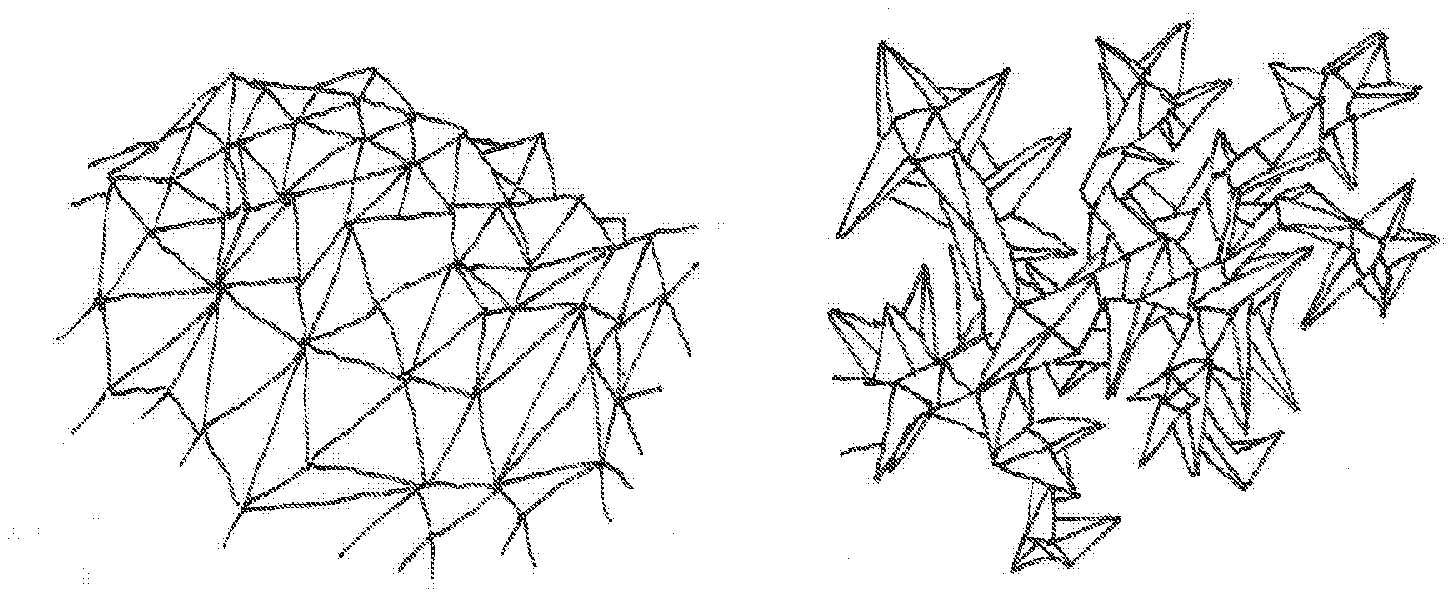}
\end{center}

\noindent{\small Figure 9.
A pictorial description of the 
smooth (left) and rough (right) phases
of four-dimensional lattice
quantum gravity.}
\medskip

\label{fig:smooth}

There are a number of ways by which the critical exponents can
be determined accurately from numerical simulations, but it is beyond
the scope of this review to go into details.
For example, one way to extract the critical exponent $\nu$ is to fit the average curvature
to the form [see Eq.~(\ref{eq:rsing})]
\beq
{\cal R} (k) \; \mathrel{\mathop\sim_{ k \rightarrow k_c}}
- A_{\cal R} \, ( k_c - k )^\delta \;\;\;\; .
\label{eq:rsing1}
\eeq
Using this general set of procedures one obtains eventually
\beq
k_c = 0.0636(11) \;\;\;\;\;  \nu = 0.335(9) \;\;\;\; ,
\eeq
which suggests $\nu = 1/3$ for pure quantum gravity.
Note that at the critical point the gravitational coupling
is not weak, $G_c \approx 0.626 $ in units of the ultraviolet
cutoff.

Often it can be advantageous to express results obtained in the cutoff
theory in terms of physical (i.e. cutoff independent) quantities.
By the latter one means quantities for which the cutoff dependence has been
re-absorbed, or restored, in the relevant definition.
As an example, an expression equivalent to Eq.~(\ref{eq:rsing}), 
relating the vacuum expectation value of the local scalar curvature 
to the physical correlation length $\xi$ , is
\beq
{ < \int d x \, \sqrt{ g } \, R(x) > \over < \int d x \, \sqrt{ g } > } 
\; \mathrel{\mathop\sim_{ G \rightarrow G_c}} \; {\rm const.} \,
\left ( l_P^2 \right )^{(d -2 - 1 / \nu )/2} \,
\left ( {1 \over \xi^2 } \right )^{ (d - 1 / \nu )/2 }
\;\; ,
\label{eq:curvature}
\eeq
which is obtained by substituting Eq.~(\ref{eq:xi-k}) 
into Eq.~(\ref{eq:rsing}).
The correct dimensions have been restored in this last equation
by supplying appropriate powers of the Planck length $l_P=G_{phys}^{1/(d-2)}$,
which involves the ultraviolet cutoff $\Lambda$.
Then for $\nu=1/3$ the result of Eq.~(\ref{eq:curvature})
becomes particularly simple
\beq
{ < \int d x \, \sqrt{ g } \, R(x) > \over < \int d x \, \sqrt{ g } > } 
\; \mathrel{\mathop\sim_{ G \rightarrow G_c}} \; 
{\rm const.} \; \, {1 \over l_P \, \xi }  
\label{eq:curvature1}
\eeq
Note that a naive estimate based on dimensional arguments would
have suggested the incorrect result $\sim 1 / l_P^2 $.
Instead the above expression actually vanishes at the critical point.
This shows that $\nu$ plays the role of an anomalous dimension,
determining the magnitude of deviations from naive dimensional
arguments.
It is amusing to note that the value $\nu=1/3$ for gravity 
does not correspond to any
known field theory or statistical mechanics model in four
dimensions. 
For a perhaps related system, namely dilute branched polymers,
it is known that $\nu=1/2$
in three dimensions, and $\nu=1/4$ at the upper critical
dimension $d=8$, so one would expect a value close to
$1/3$ somewhere in between.
On the other hand for a scalar field one would have obtained $\nu=1$ in $d=2$
and $\nu=\half$ for $d \ge 4$, which seems excluded.

\begin{table}

\begin{center}
\begin{tabular}{|l|l|l|}
\hline\hline
& & 
\\
Method & $\nu^{-1}$ in $d=3$ & $\nu^{-1}$ in $d=4$ 
\\ \hline \hline
lattice & 1.67(6) & -
\\ \hline
lattice  & - & 2.98(7) 
\\ \hline
$2+\epsilon$ & 1.6 & 4.4
\\ \hline
truncation & 1.2 & 2.666
\\ \hline \hline
exact ? & 1.5882 & 3 
\\ \hline \hline
\end{tabular}
\end{center}
\label{exp2}

{\small {\it
Table I: Direct determinations of the critical exponent $\nu^{-1}$
for quantum gravitation, using various analytical and numerical
methods in three and four space-time dimensions.}
\medskip}


\end{table}
\vskip 10pt

Table I provides a summary of the critical
exponents for quantum gravitation as obtained by various perturbative
and non-perturbative methods in three and four dimensions.
The $2+\epsilon$ and the truncation method results were discussed
previously.
The lattice model of Eq.~(\ref{eq:zlatt}) in four dimensions
gives for the critical point $G_c \approx 0.626 $ in units of
the ultraviolet cutoff, and $ \nu^{-1} = 2.98(7)$ 
which is used for comparison in Table I.
In three dimensions the numerical results are consistent with the universality
class of the interacting scalar field.

\section{Renormalization Group and Lattice Continuum Limit}
\label{sec:contlim}

The discussion in the previous sections points to the existence
of a phase transition in the lattice gravity theory, with divergent
correlation length in the vicinity of the critical point,
as in Eq.~(\ref{eq:xi-k})
\beq
\xi (k) \; \mathrel{\mathop\sim_{ k \rightarrow k_c}} \; A_\xi \;
| k_c - k | ^{ -\nu }
\label{eq:xi-k1}
\eeq
One expects the scaling result of Eq.~(\ref{eq:xi-k1})
close to the fixed point, which we choose to rewrite here
in terms of the inverse correlation length $m \equiv 1 / \xi$
\beq
m \, = \, \Lambda \, A_m \, | \, k \, - \, k_c \, |^{ \nu } \;\; .
\label{eq:m-latt}
\eeq
In the above expression the correct dimension for $m$ (inverse length)
has been restored by inserting explicitly on the r.h.s. the ultraviolet
cutoff $\Lambda$.
Here $k$ and $k_c$ are of course still dimensionless quantities, and
correspond to the bare microscopic couplings at the
cutoff scale, $k \equiv k (\Lambda) \equiv 1/( 8 \pi G(\Lambda) )$.
$A_m$ is a calculable numerical constant, related to $A_\xi$ in 
Eq.~(\ref{eq:xi-k}) by $A_m = A_\xi^{-1}$.
It is worth pointing out that the above expression for $m (k) $ is almost identical in structure
to the one for the non-linear $\sigma$-model in the $2+\epsilon$
expansion, Eq.~(\ref{eq:m-sigma}) and in the
large $N$ limit.
It is of course also quite similar to $2+\epsilon$ result for
continuum gravity, Eq.~(\ref{eq:m-cont1}).

The lattice continuum limit corresponds to the large cutoff limit
taken at {\it fixed} $m$ or $\xi$,
\beq
\Lambda \rightarrow \infty \; , 
\;\;\;\; k \rightarrow k_c \; ,
\;\;\;\; m \; {\rm fixed} \; ,
\eeq
which shows that the continuum limit is reached in the
vicinity of the ultraviolet fixed point
(see Fig.10.).
Phrased equivalently, one takes the limit in which the
lattice spacing $a \approx 1/ \Lambda$ is sent to
zero at {\it fixed} $\xi=1/m$, which
requires an approach to the non-trivial UV fixed
point $ k \rightarrow k_c$. 
The quantity $\xi$ is supposed to be a renormalization group
invariant, a physical scale independent of the scale
at which the theory is probed.
In practice, since the cutoff ultimately determines
the physical value of Newton's constant $G$, $\Lambda$
cannot be taken to $\infty$.
Instead a very large value will suffice, $\Lambda^{-1} \sim 10^{-33} cm$,
for which it will still be true that $ \xi \gg \Lambda$ which
is all that is required for the continuum limit.

For discussing the renormalization group behavior
of the coupling it will be more convenient
to write the result of Eq.~(\ref{eq:m-latt})
directly in terms of Newton's constant $G$ as
\beq
m \, = \,  \Lambda \, \left ( { 1 \over a_0 } \right )^\nu \, 
\left [ { G ( \Lambda ) \over G_c } - 1 \right ]^ \nu 
\; ,
\label{eq:m-latt1}
\eeq
with the dimensionless constant $a_0$ related to
$A_m$ by $A_m = 1 / (a_0 k_c)^\nu $.
Note that the above expression only involves the dimensionless
ratio $G(\Lambda)/G_c$, which is the only relevant
quantity here.
From the knowledge of the dimensionless constant 
$A_m$ in Eq.~(\ref{eq:m-latt})
one can estimate from first principles the value of $a_0$ in 
Eqs.~(\ref{eq:grun-latt}).
Lattice results for the correlation functions at fixed geodesic distance
give a value for $A_m \approx 0.72 $ with a significant uncertainty,
which, when combined with the values $k_c \simeq 0.0636 $ and 
$ \nu \simeq 0.335$ given above, gives 
$a_0 = 1 /( k_c \, A_m^{1/\nu}) \simeq 42$.

\medskip

\begin{center}
\epsfxsize=10cm
\epsfbox{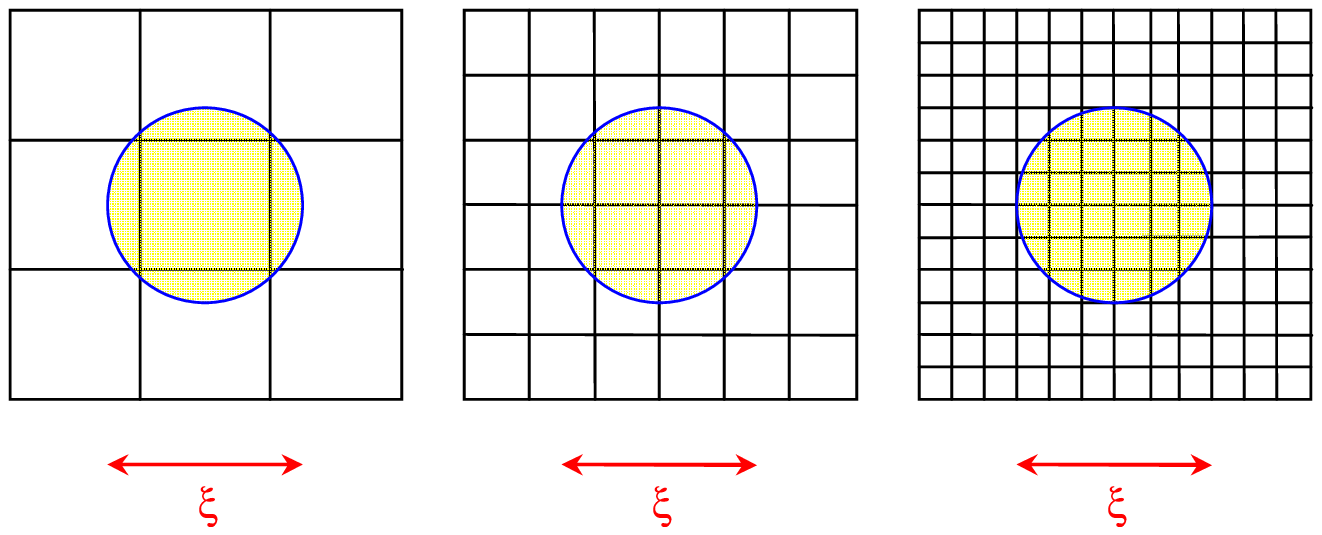 }
\end{center}

\noindent{\small Figure 10.
The lattice quantum continuum limit is gradually 
approached by
considering sequences of lattices with increasingly larger
correlation lengths $\xi$ in lattice units.
Such a limit requires the
existence of an ultraviolet fixed point, where quantum field 
correlations extend over many lattice spacing.}
\medskip

\label{fig:continuum}

The renormalization group invariance of the physical quantity
$m$ requires that the running gravitational
coupling $G(\mu)$ varies in the vicinity of the
fixed point in accordance with the above equation, with
$\Lambda \rightarrow \mu$, where $\mu$ is now an arbitrary scale,
\beq
m \, = \,  \mu \, \left ( { 1 \over a_0 } \right )^\nu \,
\left [ { G ( \mu ) \over G_c } - 1 \right ]^ \nu 
\; ,
\label{eq:m-mu}
\eeq
The latter is equivalent to the renormalization group
invariance requirement
\beq
\mu \, { d \over d \, \mu } \, m ( \mu , G( \mu ) ) \, = \, 0
\label{eq:m-rg-latt}
\eeq
provided $G(\mu)$ is varied in a specific way.
Eq.~(\ref{eq:m-rg-latt}) can therefore be used to obtain, if one so wishes,
$a$ $\beta$-function for the coupling $G(\mu)$ in units of the ultraviolet cutoff,
\beq
\mu \, { \partial \over \partial \, \mu } \, G ( \mu ) \; = \; 
\beta ( G ( \mu ) ) \;\; ,
\label{eq:beta-g-mu}
\eeq
with $\beta (G)$ given in the vicinity of the non-trivial fixed point,
using Eq.~(\ref{eq:m-mu}), by
\beq
\beta (G ) \, \equiv \, 
\mu \, { \partial \over \partial \, \mu } \, G( \mu )
\; \mathrel{\mathop\sim_{ G \rightarrow G_c}} \; 
- \, { 1 \over \nu } \, ( G- G_c ) + \dots \;\; .
\label{eq:beta-g-latt}
\eeq
The above procedure is in fact in complete analogy to what is done
for the non-linear $\sigma$-model.
But the last two steps are not really necessary, for
one can obtain the scale dependendence of the gravitational
coupling directly from Eq.~(\ref{eq:m-mu}), by simply solving for
$G(\mu)$,
\beq
G( \mu ) \; = \;  
G_c \left [ 1 \, + \, a_0 (m^2 / \mu^2 )^{1 / 2 \nu } + \dots \right ]
\label{eq:grun-latt} 
\eeq
This last expression can be compared directly to the $2+\epsilon$
result of Eq.~(\ref{eq:grun-cont1}), as well as to
the $\sigma$-model result of Eq.~(\ref{eq:grun-nonlin}).
The physical dimensions of $G$ can be restored by multiplying the
above expression on both sides by the ultraviolet cutoff $\Lambda$,
if one so desires.
One concludes that the above result physically implies gravitational
anti-screening: the gravitational coupling $G$ increases with distance.
In conclusion, the lattice result for $G(\mu)$ in Eq.~(\ref{eq:grun-latt})
and the $\beta$-function in Eq.~(\ref{eq:beta-g-latt}) are qualitatively
similar to what one finds both in the $2+\epsilon$ expansion
for gravity and in the non-linear $\sigma$-model {\it in the strong coupling
phase}.

\section{Curvature Scales and Gravitational Condensate}
\label{sec:curvature}

As can be seen from Eq.~(\ref{eq:rescale}) the path
integral for pure quantum gravity can be made to depend on the gravitational
coupling $G$ and the cutoff $\Lambda$ only:
by a suitable rescaling of the metric, or the edge lengths in the discrete case,
one can set the cosmological constant to unity in units of the cutoff.
The remaining coupling $G$ should then be viewed more appropriately
as the gravitational constant
{\it in units of the cosmological constant $\lambda$}.

The renormalization group running of $G (\mu)$ in Eq.~(\ref{eq:grun-latt})
involves an invariant scale $\xi=1/m$.
At first it would seem that this scale could take any value, including
very small ones based on the naive estimate $\xi \sim l_P$, which would
preclude any observable quantum effects in the foreseeable future.
But the result of Eqs.~(\ref{eq:curvature}) and (\ref{eq:curvature1}) suggest
otherwise, namely that the non-perturbative scale $\xi$ is in fact related
to {\it curvature}.
From astrophysical observation the average curvature is very small, 
so one would conclude from Eq.~(\ref{eq:curvature1}) that $\xi$ is very
large, and possibly macroscopic.
But the problem with Eq.~(\ref{eq:curvature1}) is that it involves the 
lattice Ricci scalar, a quantity related curvature probed by parallel
transporting vectors around infinitesimal loops with size comparable
to the lattice cutoff $\Lambda^{-1}$.
What one would like is instead a relationship between $\xi$ and
quantities which describe the geometry on larger scales.

In many ways the quantity $m$ of Eq.~(\ref{eq:m-mu}) behaves as a 
dynamically generated mass scale, quite similar to what
happens in the non-linear $\sigma$-model case,
or in the $2+\epsilon$ gravity case [Eq.~(\ref{eq:m-cont})].
From the classical field equation $R=4 \lambda$ one can relate the above $\lambda$, and therefore the mass-like parameter $m$,
to curvature, which leads to the identification
\beq
\lambda_{obs} \; \simeq \; { 1 \over \xi^2 } 
\label{eq:xi_lambda}
\eeq 
with $\lambda_{obs}$ the observed small but non-vanishing cosmological constant.

A further indication that the identification of the observed scaled cosmological
constant with a mass-like - and thefore renormalization group invariant - term 
makes sense beyond the weak field limit can be seen for example
by comparing the structure of the three classical field equations
\bea
R_{\mu\nu} \, - \, \half \, g_{\mu\nu} \, R \, + \, \lambda \, g_{\mu\nu} \; 
& = & \; 8 \pi G  \, T_{\mu\nu}
\nonumber \\
\partial^{\mu} F_{\mu\nu} \, + \, \mu^2 \, A_\nu \, 
& = & \; 4 \pi e \, j_{\nu} 
\nonumber \\
\partial^{\mu} \partial_{\mu} \, \phi \, + \, m^2 \, \phi \; 
& = & \; {g \over 3!} \, \phi^3
\label{eq:masses}
\eea
for gravity, QED (massive via the Higgs mechanism) and a self-interacting scalar
field, respectively.

A third argument suggesting the identification of the scale $\xi$
with large scale curvature, and therefore with the observed scaled 
cosmological constant, goes as follows.
Observationally the curvature on large scale can be determined by
parallel transporting vectors around very large loops,
with typical size much larger than the lattice cutoff $l_P$.
In gravity, curvature is detected by parallel transporting vectors around
closed loops.
This requires the calculation of a path dependent product of
Lorentz rotations ${\bf R}$, in the Euclidean case elements of $SO(4)$,
as discussed earlier.
From it then follows the identification of 
the correlation length $\xi$ with a measure of large scale curvature,
the most natural candidate being the scaled cosmological constant
$\lambda_{phys} $, as in Eq.~(\ref{eq:xi_lambda}).
This relationship, taken at face value, implies a very large, cosmological value
for $\xi \sim 10^{28} cm$, given the present bounds on $\lambda_{phys}$.
Thus the modified Einstein equations, incorporating the
quantum running of $G$, should read
\beq
R_{\mu\nu} \, - \, \half \, g_{\mu\nu} \, R \, + \, \lambda \, g_{\mu\nu}
\; = \; 8 \pi \, G(\mu)  \, T_{\mu\nu}
\label{eq:field0}
\eeq
with $\lambda \simeq { 1 \over \xi^2 } $.
Here only $G(\mu)$ on the r.h.s. scale-dependent in accordance with
Eq.~ (\ref{eq:grun-latt}).
The precise meaning of $G(\mu)$ in a covariant framework
will be given shortly.


It is worth pointing out here that the gravitational vacuum condensate,
which only exists in the strong coupling phase $G>G_c$, and which is
proportional to the curvature, is genuinely non-perturbative.
Thus one can summarize the result of Eq.~(\ref{eq:xi_lambda}) as
\beq
{\cal R }_{obs} \; \simeq \; (10^{-30} eV)^2 \, \sim \, \xi^{-2} 
\eeq
where the condensate is, according to Eq.~(\ref{eq:m-latt1}),
non-analytic at $G=G_c$.
A graviton vacuum condensate of order $\xi^{-1} \sim 10^{-30} eV$ is
of course extraordinarily small compared to the QCD color condensate 
($\Lambda_{\overline{MS}} \simeq 220 \, MeV$) and the electro-weak Higgs condensate ($v \simeq 250 \, GeV$).
One can pursue the analogy with non-Abelian gauge theories further
by stating that the quantum gravity theory cannot provide a value for the 
non-perturbative curvature scale $\xi$:
it needs to be fixed by some sort of phenomenological input, either by
Eq.~(\ref{eq:grun-latt}) or by Eq.~(\ref{eq:xi_lambda}).
But one important message is that the scale $\xi$ in those two equations is
one and the same.

\section{Effective Field Equations }
\label{sec:effective}

To summarize the results of the previous section,
the result of Eq.~(\ref{eq:grun-latt}) implies
for the running gravitational coupling in the vicinity of the
ultraviolet fixed point
\beq
G(k^2) \; = \; G_c \left [ \; 1 \, 
+ \, a_0 \left ( { m^2 \over k^2 } \right )^{1 \over 2 \nu} \, 
+ \, O ( \, ( m^2 / k^2 )^{1 \over \nu} ) \; \right ]
\label{eq:grun-k}
\eeq
with $m=1/\xi$, $a_0 > 0$ and $\nu \simeq 1/3$.
Since $\xi$ is expected to be very large,
the quantity $G_c$ in the above expression should now
be identified with the laboratory scale value 
$ \sqrt{G_c} \sim 1.6 \times 10^{-33} cm$.
The effective interaction in real space is often obtained by Fourier transform,
but the above expression is a bit singular as $k^2 \rightarrow 0$.
The infrared divergence needs to be regulated, which can be achieved
by utilizing as the lower limit of momentum integration
$m=1/\xi$.
Alternatively, as a properly infrared regulated version of the above
expression one can use
\beq
G(k^2) \; \simeq \; G_c \left [ \; 1 \, 
+ \, a_0 \left ( { m^2 \over k^2 \, + \, m^2 } \right )^{1 \over 2 \nu} \, 
+ \, \dots \; \right ]
\label{eq:grun-k-reg}
\eeq
Then at very large distances $r \gg \xi$  the gravitational coupling approaches
the finite value $G_\infty = ( 1 + a_0 + \dots ) \, G_c $.

The first step in analyzing the consequences of a running of $G$
is to re-write the expression for $G(k^2)$ in a coordinate-independent
way, for example by the use of a non-local Vilkovisky-type effective actions.
Since in going from momentum to position space one 
usually employs $k^2 \rightarrow - \Box$,
to obtain a quantum-mechanical running of the gravitational
coupling one needs to make the replacement
\beq
G  \;\; \rightarrow \;\; G( \Box )
\label{eq:gbox}
\eeq
and therefore from Eq.~(\ref{eq:grun-k})
\beq
G( \Box ) \, = \, G_c \left [ \; 1 \, 
+ \, a_0 \left ( { 1\over \xi^2 \Box  } \right )^{1 \over 2 \nu} \, 
+ \, \dots \, \right ] \; .
\label{eq:grun-box}
\eeq
The running of $G$ is expected to lead to
a non-local gravitational action, for example of the form
\beq
I \; = \; { 1 \over 16 \pi G } \int dx \sqrt{g} \,
\left ( 1 \, - \, a_0 \, 
\left [ {1 \over \xi^2 \Box } \right )^{ 1 / 2 \nu} \, 
+ \dots \right ] \, R \; \; .
\label{eq:ieff_sr}
\eeq
Due to the fractional exponent in general the covariant operator appearing in the above expression, namely
\beq
A (\Box) \; = \; a_0
\left( { 1 \over \xi^2 \Box } \right)^{1/2\nu} 
\label{eq:abox}
\eeq
has to be suitably defined by analytic continuation from positive
integer powers.
The latter can be done, for example, by computing $\Box^n$ for positive
integer $n$ and then analytically continuing to $n \rightarrow -1/2\nu$.

Had one {\it not} considered the action of Eq.~(\ref{eq:ieff_sr})
as a starting point for
constructing the effective theory, one would naturally be led 
(following Eq.~(\ref{eq:gbox}))
to consider the following effective field equations
\beq
R_{\mu\nu} \, - \, \half \, g_{\mu\nu} \, R \, + \, \lambda \, g_{\mu\nu}
\; = \; 8 \pi G  \, \left( 1 + A( \Box ) \right) \, T_{\mu\nu}
\label{eq:field1}
\eeq
the argument again being the replacement 
$G \, \rightarrow \, G(\Box) \equiv G \left( 1 + A( \Box ) \right)$.
Being manifestly covariant, these expressions at least satisfy some
of the requirements for a set of consistent field equations
incorporating the running of $G$.
The above effective field equation can in fact be re-cast in a form
similar to the classical field equations
\beq
R_{\mu\nu} \, - \, \half \, g_{\mu\nu} \, R \, + \, \lambda \, g_{\mu\nu}
\; = \; 8 \pi G  \, {\tilde T_{\mu\nu}}
\eeq
with $ {\tilde T_{\mu\nu}} \, = \, \left( 1 + A( \Box ) \right) \, T_{\mu\nu}$
defined as an effective, or gravitationally dressed, energy-momentum tensor.
Just like the ordinary Einstein gravity case,
in general ${\tilde T_{\mu\nu}}$ might not be covariantly conserved a priori,
$\nabla^\mu \, {\tilde T_{\mu\nu}} \, \neq \, 0 $, but ultimately the
consistency of the effective field equations demands that it
be exactly conserved, in consideration of the Bianchi identity satisfied
by the Riemann tensor.
In a sense the running of $G$ can be interpreted as due to some sort of 
"vacuum fluid", introduced to account for the vacuum polarization
contribution, whose energy momentum tensor one would expect to be 
ultimately covariantly conserved.
That the procedure is consistent in general is not entirely clear, in which case
the present approach should perhaps be limited to phenomenological considerations.

\section{Static Isotropic Solution}
\label{sec:static}

One can show that the quantum correction due
to the running of $G$ can be described, at least in the non-relativistic
limit of Eq.~(\ref{eq:grun-k-reg}) as applied to Poisson's equation,
in terms of a vacuum energy density $\rho_m(r)$, distributed around
the static source of strength $M$ in accordance with the result
\beq
\rho_m (r) \; = \; { 1 \over 8 \pi } \, c_{\nu} \, a_0 \, M \, m^3 \,
( m \, r )^{ - {1 \over 2} (3 - {1 \over \nu}) }  
\, K_{ {1 \over 2} ( 3 - {1 \over \nu} ) } ( m \, r ) 
\label{eq:rho_vac}
\eeq
with a constant
\beq
c_{\nu} \; \equiv \; { 2^{ {1 \over 2} (5 - {1 \over \nu}) }
\over \sqrt{\pi} \, \Gamma( {1 \over 2 \, \nu} ) } \;\; .
\label{eq:rho_vac1}
\eeq
and such that 
\beq
4 \, \pi \, \int_0^\infty \, r^2 \, d r \, \rho_m (r) \; = \;  a_0 \, M  \;\; .
\label{eq:rho_vac2}
\eeq
In the relativistic context, a manifestly covariant implementation of the 
running of $G$, 
via the $G(\Box)$ given in Eq.~(\ref{eq:grun-box}),
will induce a non-vanishing effective pressure term.
It is natural therefore to attempt to represent the vacuum polarization cloud
by a relativistic perfect fluid, with energy-momentum tensor
$ T_{\mu\nu} \;  = \; \left ( \, p \, + \, \rho \, \right ) \, u_\mu \, u_\nu
\, + \, g_{\mu \nu} \, p $.
Solving the resulting field equations gives a solution only for $\nu=1/3$ and
one finds
\beq
A^{-1} (r) \; = \; = \; B (r) \; = \; 1 \, - { 2 \, M \, G \over r } \, + \, 
{4 \, a_0 \, M \, G \, m^3 \over 3 \, \pi } \, r^2 \, \ln \, ( m \, r ) 
\, + \, \dots
\label{eq:a_small_r3}
\eeq
After a bit of work one can then obtain an expression for the
effective pressure $p_m(r)$, and one finds again in the limit $r \gg 2 M G $ 
\beq
p_m (r) \; = \; {a_0 \over 2 \pi^2 } \, M \, m^3 \, \ln \, (  m \, r ) 
\, + \, \dots
\label{eq:p_vac_3}
\eeq
The expressions for $A(r)$ and $B(r)$ are therefore consistent with a gradual slow increase 
in $G$ with distance, in accordance with the formula
\beq
G \; \rightarrow \; G(r) \; = \; 
G \, \left ( 1 \, + \, 
{ a_0 \over 3 \, \pi } \, m^3 \, r^3 \, \ln \, { 1 \over  m^2 \, r^2 }  
\, + \, \dots
\right )
\label{eq:g_small_r3}
\eeq
in the regime $r \gg 2 \, M \, G$,
and therefore of course in agreement with the original result of 
Eqs.~(\ref{eq:grun-k})
or (\ref{eq:grun-k-reg}), namely that the classical laboratory value of $G$ 
is obtained for $ r \ll \xi $.
There are similarities, as well as some rather substantial differences,
with the corresponding QED small $r$ result
\beq
Q \; \rightarrow \; Q(r) \; = \; Q \, \left ( 1 \, + \, 
{\alpha \over 3 \, \pi } \, \ln { 1 \over m^2 \, r^2 } \, + \, \dots
\right )
\label{eq:qed_s}
\eeq
In the gravity case, the correction vanishes as $r$ goes to zero: in this
limit one is probing the bare mass, unobstructed by its virtual graviton cloud.
In some ways the running $G$ term acts as a local cosmological
constant term, for which the
$r$ dependence of the vacuum solution for small $r$ is fixed by the nature
of the Schwarzschild solution with a cosmological constant term.
One can therefore wonder what these solutions might look like in $d$
dimensions, and after some straightforward calculations one finds
that in $d \ge 4 $ dimensions only $\nu=1/(d-1)$ is possible.

\section{Cosmological Solutions}
\label{sec:cosm}

A scale dependent Newton's constant will lead to small modifications
of the standard cosmological solutions to the Einstein field
equations.
Here we will provide a brief discussion of what modifications are
expected from the effective field equations on the basis of $G(\Box)$,
as given in Eq.~(\ref{eq:gbox}), which itself originates in
Eqs.~(\ref{eq:grun-k-reg}) and (\ref{eq:grun-k}).
The starting point is the quantum effective field equations
of Eq.~(\ref{eq:field1}), 
\beq
R_{\mu\nu} \, - \, \half \, g_{\mu\nu} \, R \, + \, \lambda \, g_{\mu\nu}
\; = \; 8 \pi G  \, \left( 1 + A( \Box ) \right) \, T_{\mu\nu}
\label{field2}
\eeq
with $A(\Box)$ defined in Eq.~(\ref{eq:abox}).
In the Friedmann-Robertson-Walker (FRW) framework these are
applied to the standard homogenous isotropic metric
\beq
ds^2 \; = \; - dt^2 + a^2(t) \, \left \{ { dr^2 \over 1 - k\,r^2 } 
+ r^2 \, \left( d\theta^2 + \sin^2 \theta \, d\varphi^2 \right)  \right \}
\eeq
It should be noted that there are {\it two} quantum contributions to the
above set of effective field equations. 
The first one arises because of the presence of a non-vanishing 
cosmological constant $\lambda \simeq 1 / \xi^2 $ caused by the
non-perturbative vacuum condensate of Eq.~(\ref{eq:xi_lambda}).
As in the case of standard FRW cosmology, this is expected to be 
the dominant contributions at large times $t$, and gives an exponential
(for $\lambda>0$ or cyclic (for $\lambda < 0$) expansion of the scale factor.
The second contribution arises because of the running of $G$ in the effective field equations,
\beq
G(\Box)  \; = \; G \, \left( 1 + A( \Box ) \right) \; = \; 
G \, \left [ \, 1 + a_0 \left ( \xi^2 \Box \right )^{-{ 1 \over 2 \nu }} \, + \, \dots \, \right ]
\eeq
for for $t \ll \xi$, with $\nu \simeq 1/3$ and $a_0>0 $ a calculable 
coefficient of order one [see Eqs.~(\ref{eq:grun-k}) and (\ref{eq:grun-k-reg})].
The next step is to examine the full effective field equations with a
cosmological constant $\lambda=0$,
\beq
R_{\mu\nu} \, - \, \half \, g_{\mu\nu} \, R \, 
\; = \; 8 \pi G  \, \left( 1 + A( \Box ) \right) \, T_{\mu\nu}
\eeq
Here the d'Alembertian operator
\beq
\Box \; = \; g^{\mu\nu} \nabla_\mu \nabla_\nu 
\eeq
acts on a second rank tensor,
\bea
\nabla_{\nu} T_{\alpha\beta} \, = \, \partial_\nu T_{\alpha\beta} 
- \Gamma_{\alpha\nu}^{\lambda} T_{\lambda\beta} 
- \Gamma_{\beta\nu}^{\lambda} T_{\alpha\lambda} \, \equiv \, I_{\nu\alpha\beta}
\nonumber
\eea
\beq 
\nabla_{\mu} \left( \nabla_{\nu} T_{\alpha\beta} \right)
= \, \partial_\mu I_{\nu\alpha\beta} 
- \Gamma_{\nu\mu}^{\lambda} I_{\lambda\alpha\beta} 
- \Gamma_{\alpha\mu}^{\lambda} I_{\nu\lambda\beta} 
- \Gamma_{\beta\mu}^{\lambda} I_{\nu\alpha\lambda} 
\eeq
Next one assumes again that $T_{\mu\nu}$ has the perfect fluid form, 
for which one obtains from the action of $\Box$ on $T_{\mu\nu}$
\bea
\left( \Box \, T_{\mu\nu} \right )_{tt} \; & = & \; 
6 \, \left [ \rho (t) \, + \, p(t) \right ]
\, \left ( { \dot{a}(t) \over a(t) } \right )^2
\, - \, 3 \, \dot{\rho}(t) \,  { \dot{a}(t) \over a(t) }
\, - \, \ddot{\rho}(t) 
\nonumber \\
\left( \Box \, T_{\mu\nu} \right )_{rr} \; & = & \; 
{ 1 \over 1 \, - \, k \, r^2 } \left \{
2 \, \left [ \rho (t) \, + \, p(t) \right ] \, \dot{a}(t)^2 
\, - \, 3 \, \dot{p}(t) \, a(t) \, \dot{a}(t) 
\, - \, \ddot{p}(t) \, a (t)^2  \right \}
\nonumber \\
\left( \Box \, T_{\mu\nu} \right )_{\theta\theta} \; & = & \; 
r^2 \, ( 1 \, - \, k \, r^2 ) \, 
\left( \Box \, T_{\mu\nu} \right )_{rr}
\nonumber \\
\left( \Box \, T_{\mu\nu} \right )_{\varphi\varphi} \; & = & \; 
r^2 \, ( 1 \, - \, k \, r^2 ) \, 
\sin^2 \theta \, \left( \Box \, T_{\mu\nu} \right )_{rr}
\label{eq:boxont}
\eea
with the remaining components equal to zero.
Note that a non-vanishing pressure contribution is generated in the effective
field equations, even if one assumes initially a pressureless fluid, $p(t)=0$.
As before, repeated applications of the d'Alembertian $\Box$ to the above expressions leads
to rapidly escalating complexity,
which can only be tamed by introducing some further simplifying assumptions,
such as a power law behavior for
the density, $\rho(t) = \rho_0 \, t^\beta$, and $p(t)=0$. 
After a lengthy calculation one finds for a universe filled with non-relativistic 
matter ($p$=0), the effective Friedmann equations then have the following appearance 
\bea
{ k \over a^2 (t) } \, + \,
{ \dot{a}^2 (t) \over a^2 (t) }  
& \; = \; & { 8 \pi G(t) \over 3 } \, \rho (t) \, + \, { 1 \over 3 \, \xi^2 }
\nonumber \\
& \; = \; & { 8 \pi G \over 3 } \, \left [ \,
1 \, + \, c_\xi \, ( t / \xi )^{1 / \nu} \, + \, \dots \, \right ]  \, \rho (t)
\, + \, \third \, \lambda 
\label{eq:fried_tt}
\eea
for the $tt$ field equation, and
\bea
{ k \over a^2 (t) } \, + \, { \dot{a}^2 (t) \over a^2 (t) }
\, + \, { 2 \, \ddot{a}(t) \over a(t) } 
& \; = \; & - \, { 8 \pi G \over 3 } \, \left [ \, c_\xi \, ( t / \xi )^{1 / \nu} 
\, + \, \dots \, \right ] \, \rho (t) 
\, + \, \lambda
\label{eq:fried_rr}
\eea
for the $rr$ field equation.
The running of $G$ appropriate for the Robertson-Walker metric,
and appearing explicitly in the first equation, is given by
\beq
G(t) \; = \; G \, \left [ \; 1 \, + \, c_\xi \, 
\left ( { t \over \xi } \right )^{1 / \nu} \, + \, \dots \, \right ]
\label{eq:grun_frw}
\eeq
with $c_\xi$ of the same order as $a_0$ of Eq.~(\ref{eq:grun-k}).
Note that the running of $G(t)$ induces as well an effective pressure term in the second 
($rr$) equation.
We wish to emphasize that we are {\it not} talking here about models with a
time-dependent value of $G$.
Thus, for example, the value of $G \simeq G_c$ at laboratory scales should be taken to
be constant throughout most of the evolution of the universe.

Finally it should be noted that the effective Friedmann equations of Eqs.~(\ref{eq:fried_tt}) 
and (\ref{eq:fried_rr}) also bear a superficial degree of resemblance to what
might be obtained in some scalar-tensor theories of gravity, where the
gravitational Lagrangian is postulated to be some singular function of the scalar curvature. 
The former scenario would then correspond the to an effective gravitational action
\beq
I_{eff}  \; \simeq \; { 1 \over 16 \pi G } \int dx \, \sqrt{g} \,
\left ( \, R \, + \, 
{ f \, \xi^{- {1\over\nu} } \over | R |^{{1\over 2 \nu} -1} } 
\, - \, 2 \, \lambda \, \right )
\label{eq:scalar-tens}
\eeq
but with $\nu=1/3$,  $f$ a numerical constant of order one, and 
$\lambda \simeq 1 / \xi^2 $.

\vskip 20pt

{\bf Acknowledgements}

The author wishes to thank Hermann Nicolai, Stefan Theisen and the
Max Planck Institut f\" ur Gravitationsphysik (Albert-Einstein-Institut)
in Potsdam for a very warm hospitality. 
This work was supported in part by the Max 
Planck Gesellschaft zur F\" orderung der Wissenschaften.

\vfill

\newpage

\end{document}